\documentclass[usenatbib]{mnras}

\usepackage{newtxtext,newtxmath}

\usepackage[T1]{fontenc}
\usepackage{ae,aecompl}

\usepackage{graphicx}	
\usepackage{amsmath}	
\usepackage{amssymb}	
\usepackage{siunitx}
\usepackage[version=4]{mhchem}
\usepackage[autostyle=true]{csquotes}
\usepackage{subcaption}
\usepackage{xcolor}
\usepackage{footnote}

\title[]{A chemical kinetics code for modelling exoplanetary atmospheres}

\author[R. Hobbs et al.]{
Richard Hobbs,$^{1}$\thanks{E-mail: rh567@cam.ac.uk}
Oliver Shorttle,$^{1,2}$
Nikku Madhusudhan$^{1}$
and Paul Rimmer$^{2,3,4}$
\\
$^{1}$Institute of Astronomy, University of Cambridge, Cambridge, CB3 0HA, UK\\
$^{2}$Cambridge Earth Sciences, University of Cambridge, Cambridge CB2 3EQ, UK\\
$^{3}$MRC Laboratory of Molecular Biology, Cambridge, CB2 OQH, UK\\
$^{4}$Cavendish Astrophysics, University of Cambridge, Cambridge, CB3 0HE, UK\\
}

\date{Accepted XXX. Received YYY; in original form ZZZ}

\pubyear{2019}

\begin{document}
\label{firstpage}
\pagerange{\pageref{firstpage}--\pageref{lastpage}}
\maketitle

\begin{abstract}
Chemical compositions of exoplanets can provide key insights into their physical processes, and formation and evolutionary histories. Atmospheric spectroscopy provides a direct avenue to probe exoplanetary compositions. However, whether obtained in transit or thermal emission, spectroscopic observations probe limited pressure windows of planetary atmospheres and are directly sensitive to only a limited set of spectroscopically active species. It is therefore critical to have chemical models that can relate retrieved atmospheric compositions to an atmosphere's bulk physical and chemical state. To this end we present \textsc{Levi} a new chemical kinetics code for modelling exoplanetary atmospheres. \textsc{Levi} calculates the gas phase hydrogen, oxygen, carbon, and nitrogen chemistry in planetary atmospheres. Here we focus on hot gas giants. Applying \textsc{Levi}, we investigate how variations in bulk \ce{C}/\ce{O} and \ce{N}/\ce{O} affects the observable atmospheric chemistry in hot Jupiters. For typical hot Jupiters we demonstrate the strong sensitivity of molecular detections to the atmospheric \ce{C}/\ce{O}. Molecular detections are conversely less sensitive to the atmospheric \ce{N}/\ce{O} ratio, although highly super-solar \ce{N}/\ce{O} can decrease the \ce{C}/\ce{O} required for \ce{HCN} and \ce{NH3} detection. Using a new P-T profile for HD 209458b without a thermal inversion, we evaluate recently reported detection's of \ce{CO}, \ce{H2O} and \ce{HCN} in its day-side atmosphere. We find that our models are consistent with the detected species, albeit with a narrow compositional window around C/O $\sim$ 1. A \ce{C}/\ce{O} $\gtrsim$ 0.9 (1.6 times solar) was required to meet the minimum reported value for \ce{HCN}, while a \ce{C}/\ce{O} $\lesssim$ 1 (1.8 times solar) was required to fit the nominal \ce{H2O} abundance.
\end{abstract}




\section{Introduction}


\newcommand{\keyword}[1]{\textbf{#1}}
\newcommand{\tabhead}[1]{\textbf{#1}}
\newcommand{\code}[1]{\texttt{#1}}
\newcommand{\file}[1]{\texttt{\bfseries#1}}
\newcommand{\option}[1]{\texttt{\itshape#1}}


Identifying and characterising features of exoplanets is the observational cornerstone to answering some of the biggest questions in planetary astrophysics: Within the field of giant planets major questions remain unresolved regarding their formation, composition, evolution during atmospheric escape, and how significant migration is in their history (\citealt{Beaug2012,Mordasini2016,Nelson2017,Dawson2018,McGuire2018}). Both the location of where giant planets form and possible later migration through the proto-planetary disk may leave signs in the compositions of their atmospheres (e.g. \citealt{Oberg2011,Moses2013,Madhu2014,Mordasini2016}). Tracing giant planet migration history is also important for understanding terrestrial planet formation and survivability. This question is clearest for hot Jupiter systems, where the presumed inward migration of a Jupiter mass object may have significantly disrupted habitable zone planets (\citealt{Popolo2005,Hands2016,Sanchez2018}). 

A key tool for characterising an exoplanet's atmosphere is its spectrum, observed via transits or thermal emission. The molecules in a planet's atmosphere absorb and emit light at particular frequencies, which can imprint a signature of the planet's atmosphere on the light received at Earth from the planet-star system. Thus, it is possible to infer atmospheric chemistry based upon the features of the planet's transmission or emission spectrum (\citealt{Seager2000,Brown2001a,Deming2005,Grillmair2008,Seager2010,Madhu2016}).

However, spectra only contain emission or absorption features for the spectrally active species present in a planet's atmosphere, and therefore provide only a partial view of atmospheric chemistry and dynamics. In addition, the observed transmission or emission spectrum comes from a narrow band of pressure ranges, generally between $1$ - $10^{-3}$ bar, thus lacking sensitivity to chemistry in the rest of the atmosphere. Yet, with vertical mixing and diffusion transporting species through the atmosphere, chemistry at higher and lower pressures can contribute to atmospheric species in the observed pressure window. It is also possible for layers of clouds or hazes to block observations, resulting in muted spectral features that can imply incorrect abundances. This means that models are required to complete the picture of physical and chemical processes occurring in a planet.

Atmospheric models can fill in details of atmospheric composition and structure that retrievals are not directly sensitive to, enabling physically and chemically informed interpolation and extrapolation of the retrieval observations. Improving the estimates of bulk atmospheric composition from retrievals is also then possible, by comparing the model atmosphere at different \ce{C}/\ce{O} and \ce{N}/\ce{O} ratios to the retrieved atmospheric abundances. Since the composition of planetary atmospheres may reflect when and where in a disk planets formed (\citealt{Oberg2011,Madhu2014,Cleeves2018}), these elemental ratios may ultimately provide a key insight into the history of planet formation.

To construct the chemical model, a network of atmospheric chemical reactions is compiled and used as input to a model that calculates the production and loss of all species in the atmosphere. This produces a profile of chemical abundances for an atmosphere in thermochemical equilibrium. However, a planetary atmosphere may not be in chemical equilibrium over its entire extent. It is expected that the deeper, hotter parts of the atmosphere (P $\gtrsim$ 0.1 - 1 bar) approach thermochemical equilibrium, however, higher up (at lower pressures) various non-equilibrium processes can become significant. Therefore, to more accurately model atmospheric chemistry it is necessary to include the non-equilibrium processes that can dominate at lower pressures, such as vertical mixing and photo-chemistry (\citealt{Cooper2006,Moses2011,Moses2013}).

We report a new code for 1-D modelling of chemistry in exoplanetary atmospheres, named \textsc{Levi}. \footnote{Named after the Latin \textit{Levis}, meaning light.} The chemical processes that it can model include thermochemical equilibrium, eddy-diffusion, molecular diffusion, thermal diffusion and photochemistry. Through inputs in the form of pressure-temperature profiles, vertical mixing profiles, stellar properties, and a network of possible chemical reactions, the model can predict the mixing ratios of chemical species in a planet's atmosphere. Currently, the code uses a network that includes molecules comprised of \ce{H}, \ce{C}, \ce{O}, \ce{N} or \ce{He}. The code is built to be modular in the chemical processes and reactions it can model, and the species it includes. We use this functionality to better understand the effects any of these components of the model can have on the chemistry of the atmosphere. Here, we focus on applying the code to the hydrogen rich atmospheres of hot Jupiters, although it can be generalised to apply to the atmospheres of terrestrial planets.

\begin{savenotes}
\begin{table*}
\begin{tabular}{llcccccc}
\hline
\multicolumn{1}{c}{Model}             & \multicolumn{1}{c}{Network Basis}               & Diffusion & Photochemistry & Elements & Reactions & Molecules    \\ \hline
\cite{Liang2003} &     From Laboratory Measurements \footnote{These rates were measured for a low temperature Jovian Model.}                            & Yes      & Yes            & H/C/O & 253 & Unknown     \\
\cite{Zahnle2009} &               \cite{Zahnle1995}                                   & Yes      & Yes            & H/C/O/N/S & 507 & 49     \\
\cite{Moses2011}  & \cite{Gladstone1996}          & Yes      & Yes            & H/C/O/N   & 1600 & 90     \\
\cite{Venot2012} & From Combustion Studies                                     & Yes      & Yes            & H/C/O/N  &1918&104      \\
\cite{Hu2012}                                            & New         & Yes      & Yes            & H/C/O/N/S  &800&111      \\
\cite{R2016}      & Stand2015                                      & Yes      & Yes            & H/C/O/N/Metals &2980&   162 \\
\cite{Drummond2016}      & \cite{Venot2012}                                    & Yes      & Yes            & H/C/O/N  &1918&104      \\
\cite{Tsai2017}   & \cite{Glassman2015}                                             & Yes      & No             & H/C/O  &300&29        \\
This work                             & Stand2018                                         & Yes      & Yes            & H/C/O/N &2000&150       \\ \hline

\end{tabular}
\caption{A comparison of the main features of a selection of non-equilibrium atmospheric chemistry codes in the literature.}
\label{tab:comp}
\end{table*}

Several models of exoplanetary atmospheric chemistry have been previously reported in the literature. While some focus purely on thermochemical equilibrium (\citealt{Burrows1999,Lodders2002,Blecic2016}), others have also sought to encapsulate the non-equilibrium chemistry occurring (i.e., \citealt{Liang2003,Zahnle2009,Moses2011,Hu2012,Miller2012,Venot2012,R2016,Tsai2017}). Table \ref{tab:comp} shows a comparison of the processes included in a selection of chemical kinetics models. These models all work similarly, using a pressure - temperature (P-T) and eddy-diffusion ($K_{zz}$) profile as inputs to their codes to calculate abundance profiles of the atmosphere, with most of the differences between their predictions arising in their chemical network. \cite{Liang2003} first reported a photochemical model with \ce{H}-\ce{C}-\ce{O} chemistry for a select set of species in highly irradiated atmospheres. \cite{Zahnle2009} investigated the effects of photochemistry on sulfur bearing molecules in hot Jupiters, and later warm Jupiters (\citealt{Zahnle2016}). This model was also adopted and furthered in \cite{Miller2012}. \cite{Moses2011} and \cite{Venot2012} both produced codes that focused on \ce{H}-\ce{C}-\ce{N}-\ce{O} photochemistry on hot Jupiters. \cite{Agundez2014b} adapted the work of \cite{Venot2012} to report a 2-D model for hot Jupiters. \cite{Hu2012} created a model for the atmosphere of super-Earths, as well as incorporating the additional non-equilibrium effects that could occur on terrestrial planets. \cite{R2016} focused on the effects that ions produced by lightning could have on the chemistry of hot Jupiters, as well as including a selection of metals in their network. \cite{Drummond2016} reported a model that self-consistently solved for both the P-T profile and chemical abundance profiles. \cite{Tsai2017} use a reduced chemical network to reproduce the results of \cite{Moses2011} and \cite{R2016} for several hot Jupiters, and analyse the accuracy of the quenching approximation. Ultimately, all of these approaches, including the one presented in this work, represent a compromise between efficiency and complexity, with full 3D coupled chemical and dynamical simulations being required to capture all aspects of atmospheric chemistry.

We demonstrate \textsc{Levi} by applying it to two of the most studied hot Jupiters, HD 209458b and HD 189733b. HD 209458b and HD 189733b are excellent candidates for benchmarking \textsc{Levi} as they are some of the best characterised planets (e.g., \citealt{Seager2000,Moses2013,Lewis2013,Liu2013,Mayne2014,Schwarz2015,Amundsen2016,Brewer2017}), and several previous chemical kinetics models have been applied to them. Between the two planets, a preliminary parameter space can be explored, comparing the differences between hot Jupiters with and without thermal inversions, the influence of strong UV radiation, and the effects of an increase in a planet's effective temperature. In particular, HD 189733b is a very good candidate to study the effect of photochemistry, because of both the high stellar UV it receives and the relatively low atmospheric temperature. HD 209458b was thought to have a temperature inversion (\citealt{Knutson2008}), although that is now debated (\citealt{Diamond2014,Schwarz2015}). The temperature inversion is being kept in most of this work, both to match \cite{Moses2011} for benchmarking and to better investigate the effects of high temperature at high and low pressures. We do, however, also use a P-T profile for HD 209458b that lacks a thermal inversion when comparing our model to a recent work that provides evidence of the presence of number of molecular species in it's atmosphere (\citealt{Hawker2018}).

In Section \ref{sec:section2} the model and network being used for this code are described. Section \ref{Section3} contains a validation against previously constructed codes and an examination of how the abundances of species in the atmosphere vary from thermochemical equilibrium due to vertical transport and photochemistry. Section \ref{Section4} contains an exploration of the \ce{C}/\ce{O} and \ce{N}/\ce{O} parameter space. This is used to investigate how the bulk chemistry of an atmosphere can affect the mixing ratios of species within it, and to compare our model's predictions to evidence of species detected on HD 209458b. Section \ref{sec:Futurework} contains a short summary and discussion of the results of this work.
\end{savenotes}
\section{The Atmospheric Model} 

\label{sec:section2}

Our model considers a 1-D plane-parallel atmosphere in hydrostatic equilibrium, with various processes affecting the chemistry. Using a given P-T profile, considerations of hydrostatic equilibrium, and the ideal gas law, the thermodynamic parameters of the atmosphere are determined. In this manner we follow the standard approach of existing chemical models (see Table \ref{tab:comp}) in using pre-determined profiles of both the temperature and vertical mixing strength, without self-consistently solving the radiative transfer equation. 

We model chemical diffusion, photochemistry, and, as a limiting case, chemical equilibrium. Thermochemical equilibrium occurs once the rate of production and loss rate of every molecule due to chemical reactions is the same. Including diffusion, in the form of eddy, molecular or thermal diffusion, can lead to species being lifted from deep in the atmosphere into higher, cooler, layers or cause large, heavy, molecules to sink out of the upper atmosphere. Species that get lifted to higher atmosphere layers can have their abundance frozen in above thermochemical equilibrium values, since the time it takes for them to deplete back down to their equilibrium abundance is significantly longer than their resupply time from vertically mixing. This process is known as quenching. The eddy diffusion used is an approximation for atmospheric motion, so that the complex micro-physics of atmospheric mixing can be treated as a simple macroscopic process. In the upper atmosphere, high energy photons can cause some molecules to disassociate, and new species to form from the break-down products. The rate of these reactions is determined by the UV flux received by the planet.

\textsc{Levi} solves a set of coupled differential equations that govern the chemical kinetics (described in Section \ref{sec:Chem proc}) to obtain a steady-state solution for the chemical composition of the atmosphere. By dividing up the atmosphere into vertically distributed plane-parallel layers, these equations can be described in just one dimension. Each equation describes the time evolution of one species in the atmosphere due to production or loss from chemical interactions, or fluxes from other atmosphere layers. Other factors that could influence the evolution of the atmosphere, such as evaporation, atmospheric escape and geological outgassing are not currently implemented but, in the context of giant planets, these processes are in general not operating and/or are not chemically significant (\citealt{Moses2011}).

A steady-state solution to these equations is found by time-stepping the code using a second-order Rosenbrock solver (\citealt{Rosenbrock1963}), from an initial guess for the chemical abundances, until the production and loss rates of every species in every atmosphere layer are balanced. For a closed system this will produce a solution that is stable over large time periods.   

The code is designed to be highly configurable, allowing modelling of a wide range of exoplanets. This includes atmospheric parameters such as elemental composition, the pressure-temperature (P-T) profile of the atmosphere and the eddy diffusion coefficient. The stellar characteristics, such as the stellar spectrum and the planet-star separation, and the physical characteristics of the planet such as its surface gravity can also be changed as required to model different examples of hot Jupiters. The gravitational acceleration used throughout the atmosphere is an approximation of the average gravity, taken at the surface of the planet. For Jupiter-like planets, the `surface' is taken to be at $\mathrm{1\,bar}$. It is also possible to selectively exclude a subset of reactions, such as those that consider the effects of photochemistry, or the reactions of ions, to better understand the effects those reactions can have on the atmosphere. In addition, any reactions containing an undesired chemical species can be easily stripped away, to both further explore the atmospheric response to the presence of particular species, and to improve the computational runtime.


\subsection{Model Setup}

Here we describe the equations that govern the model being used, as well as the chosen boundary and initial conditions for the model.

\subsubsection{Chemical Processes} \label{sec:Chem proc}
The model used considers contribution from multiple processes, which can be described by a series of coupled one-dimensional continuity equations,
\begin{equation}
\frac{\partial n_{i}}{\partial t} = \mathcal{P}_{i} - \mathcal{L}_{i} - \frac{\partial \Phi_{i}}{\partial z},
\label{eq:continuity}
\end{equation}
where $n_{i}$ (\si{\per\metre\cubed}) is the number density of species $i$, with $i = 1,...,N_{i}$, with $N_{i}$ being the total number of species. $\mathcal{P}_{i}$ (\si{\per\metre\cubed\per\second}) and $\mathcal{L}_{i}$ (\si{\per\metre\cubed\per\second}) are the production and loss rates of the species $i$. $\partial t$ (\si{\second}) and $\partial z$ (\si{\metre}) are the infinitesimal time step and altitude step respectively. $\Phi_{i}$ (\si{\per\metre\squared\per\second}) is the upward vertical flux of the species, given by,
\begin{equation}
\Phi_{i} = -(K_{zz}+D_i)n_{t}\frac{\partial X_{i}}{\partial z} + D_i n_i\left(\frac{1}{H_0} - \frac{1}{H} - \frac{\alpha_{T,i}}{T}\frac{dT}{dz}\right),
\label{eq:diffusion}
\end{equation}
where $X_{i}$ and $n_{t}$ (\si{\per\metre\cubed}) are the mixing ratio and total number density of molecules such that $n_{i} = X_{i}n_{t}$. The eddy-diffusion coefficient, $K_{zz}$ (\si{\metre\squared\per\second}), approximates the rate of vertical transport and $D_i$ (\si{\metre\squared\per\second}) is the molecular diffusion coefficient of species i. $H_0$ (\si{\metre}) is the mean scale height, $H$ (\si{\metre}) is the molecular scale height, $T$ (K) is the temperature, and $\alpha_{T,i}$ is the thermal diffusion factor. The equation for the molecular diffusion coefficient is adopted from \cite{Chapman1970}, where it is defined as,
\begin{equation}
D_i = \frac{3}{8n_{t}(\frac{1}{2}[d_i + d_t])^2}\left(\frac{k_bT(m_i+m_t)}{2\pi m_i m_t}\right)^{1/2},
\end{equation}
where $m_i$ (kg) and $d_i$ (m) are the mass and diameter of species $i$, respectively, and $m_t$ (kg) and $d_t$ (m) are the average mass and average diameter of all species in the atmosphere, respectively. The thermal diffusion factor is estimated from experimental and theoretical data from \cite{Chapman1970}: for atomic hydrogen, $\alpha_{T,i} \approx -0.1(1-n_i/n_{t})$; for \ce{He}, $\alpha_{T,i} \approx 0.145(1-n_i/n_{t})$; for \ce{H2}, $\alpha_{T,i} \approx -0.38$; and for all other molecules $\alpha_{T,i} \approx 0.25$. Under the assumption that the atmosphere can be approximated to an ideal gas, the total number density, $n_{t}$ (\si{\per\metre\cubed}), is
\begin{equation}
n_{t} = \frac{P(z)}{k_b\ T(z)},
\end{equation}
where $P$ (\si{\Pa}) and $T$ (\si{\K}) are the altitude dependent pressure and temperature, and $k_b$ is the Boltzmann constant.

The production and loss rates of species $i$ can be calculated by considering all the reactions which include said species, (e.g. $n_i \ce{+} n_{i'} \rightleftharpoons n_{i''} \ce{+} n_{i'''} $), then finding the rate at which $n_i$ is produced ($k_2 n_{i''} n_{i'''}$) and lost ($k_1 n_i n_{i'}$) in each reaction, where $k_1$ and $k_2$ are the rate of the forward and reverse reactions respectively, before summing over all such reactions such that,
\begin{equation}
\mathcal{P}_{i} - \mathcal{L}_{i} = \sum (k_2 n_{i'''} n_{i''} - k_1 n_i n_{i'}).
\label{eq:prodloss}
\end{equation}

\subsubsection{Atmospheric Profile}
\label{sec:Atmos}
\begin{figure*}
    \centering
    \begin{subfigure}[t]{0.49\textwidth}
        \centering
        \includegraphics[height=2.78in]{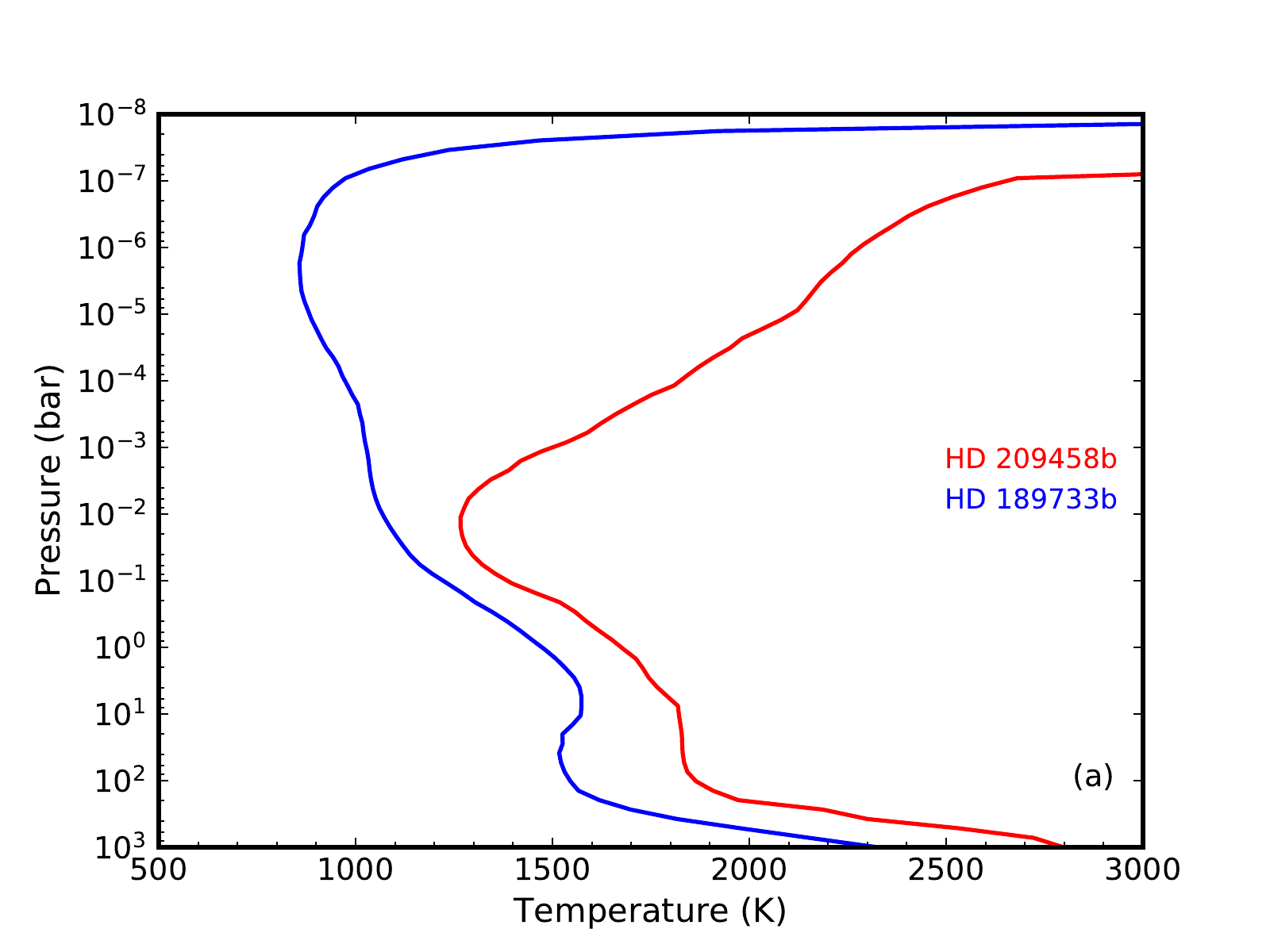}
        \caption{}
        \label{fig:PT}
    \end{subfigure}%
    \begin{subfigure}[t]{0.49\textwidth}
    	\centering
        \includegraphics[height=2.78in]{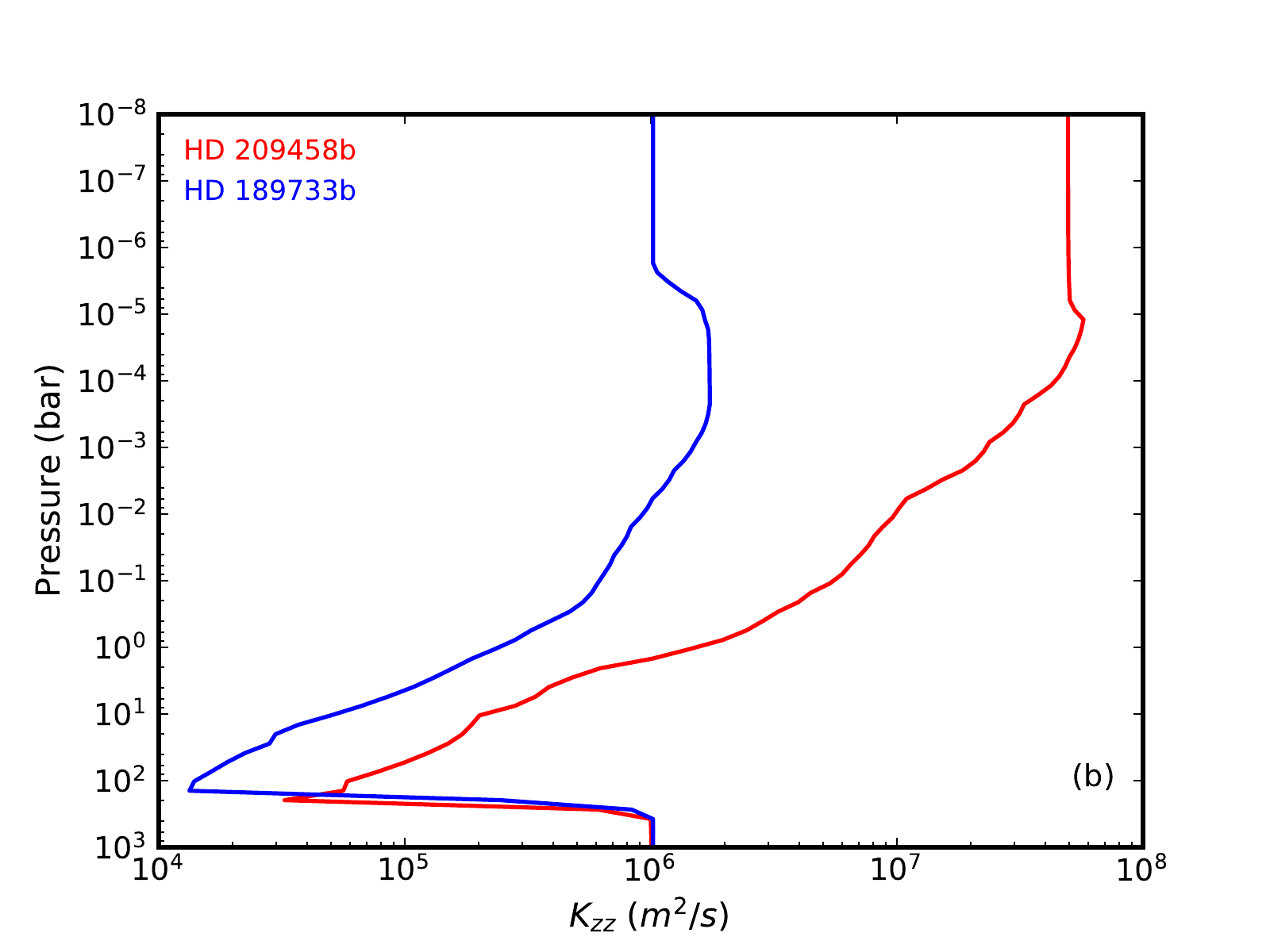}
        \caption{}
        \label{fig:Kzzprof}
    \end{subfigure}
    
    \caption[Atmospheric profiles for HD 209458b and HD 189733b]{Plot (a) are the Pressure-Temperature profiles for the average day-side of HD 209458b and HD 189733b assuming a solar composition atmosphere, adopted from \cite{Moses2011}. At pressures below $10^{-7}$ bar an approximation for the thermosphere is used from \cite{GarciaMunoz2007}. Plot (b) displays the $K_{zz}$ profiles that are being used for each planet, again drawn from \cite{Moses2011}.}
\end{figure*}

The physical characteristics of the atmospheres \textsc{Levi} simulates have a profound impact on their chemistry. This is realised through a pressure-temperature (P-T) profile that describes the atmosphere being investigated. Currently, the P-T profiles are not calculated self-consistently, but instead use pre-calculated profiles as an input. This simplification is typical for chemical kinetics models, since implementing radiative transfer to allow iterative recalculation of the P-T profile would significantly increase the runtime, in addition to the possibility of converging to a non-physical P-T profile and atmosphere chemistry. There have been other works investigating how having a self-consistent P-T profile may affect the results of chemical kinetics models. \cite{Drummond2016} found that for strong disequilibrium chemistry, self-consistent P-T profiles can vary by up to 100K. For the two test cases that we study here, the hot Jupiters HD 189733b and HD 209458b, the P-T profiles are given in Figure \ref{fig:PT}. We also consider an isothermal profile for reference.

The P-T and $K_{zz}$ profiles for HD 209458b and HD 189733b that are used in most of this work have been adopted from \cite{Moses2011}. The deepest layer of the atmosphere is chosen to be $\mathrm{10^{3}\,bar}$, a pressure at which all species are expected to be in chemical equilibrium, while the highest atmosphere layer ($\mathrm{10^{-8}\,bar}$) should be sufficient to cover the photochemical destruction and production region for all relevant molecules. For pressures below $10^{-7}$ bars, an approximation for the thermosphere is used based on the work of \cite{GarciaMunoz2007}. 

\subsubsection{Actinic Flux} \label{sec:Flux}
Many species in the atmosphere, upon absorbing a sufficiently energetic photon, can be photo-dissociated into a variety of species, including radicals. Since the rate of photo-dissociation is proportional to the number density of UV photons, it is necessary to define the actinic flux, the spherically integrated flux in any atmosphere layer due to irradiation from the host star. This can be calculated as,
\begin{equation}
F(\lambda,z) = F_0(\lambda)\ e^{-\tau(\lambda,z)/\mu_0},
\end{equation}
where $F_0$ (photons \si{\per\metre\squared\per\second} $\mathrm{nm^{-1}}$) is the flux at the top of the atmosphere (where by definition $\tau = 0$), $\mu_0$ is angle of the path of sunlight, chosen to be $57.3^o$ (\citealt{Hu2012}). $\tau$ is the total optical depth, as a function of wavelength, integrated over all the atmosphere that the light passes through, defined as,
\begin{equation}
\tau(\lambda,z) = \int_{z}^{\infty} n_i(z') \sigma_i(\lambda) dz' ,
\end{equation}
where $\sigma_i$ is the wavelength dependent cross-section of species. $\sigma_i$ is found from the sum of the molecular absorption cross-section, $\sigma_a(\lambda)$, and the Rayleigh scattering cross section (\citealt{Liou2002}),
\begin{equation}
\sigma_r(\lambda) = C_r \frac{8\pi^3(m_r(\lambda)^2-1)^2}{3\lambda^4N_s},
\end{equation}
where $m_r$ is the real part of the refractive index of the molecule, and $C_r$ is a corrective factor due to molecular anisotropy. $N_s$ is the number density at STP (1 atm, 273.15 K). For most species $C_r$ is approximately unity, apart from \ce{CO2}, for which $C_r$ is approximately 1.14 (\citealt{Sneep2005}). The refractive index can be approximated as that of the dominant species in the atmosphere, which for the atmospheres modelled in this work is \ce{H2}.

\begin{figure}
        \centering
        \includegraphics[width=0.5\textwidth]{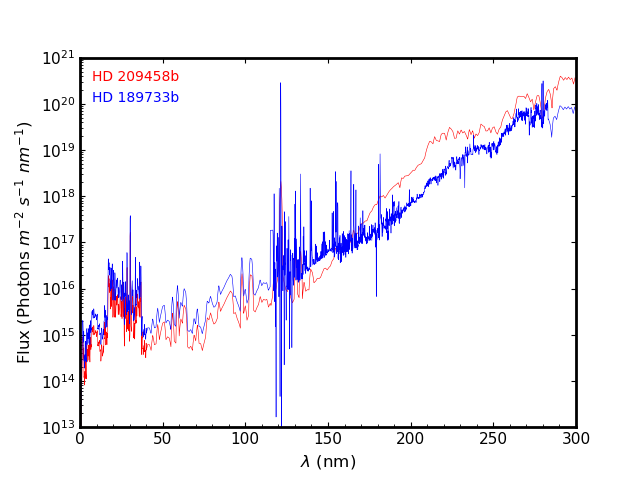}
    \caption[Actinic Fluxes]{The unattenuated UV fluxes applied to the top of the atmospheres of HD 189733b and HD 209458b. For HD 189733b the solar flux is used up to 115 nm and epsilon Eridani beyond that. For HD 209458b the solar flux is used. In both cases, the flux has been corrected to account for planet-star separation.}
     \label{fig:Flux}
\end{figure}

The host star of HD 209458b is type G0V, like the sun, and therefore the spectrum used was the unattenuated solar flux at the top of the Earth's atmosphere, weighted by the relative flux received due to differing orbital radius between Earth and the sun ($d_{\oplus}$) and the planet and its host star ($d_r$), $(d_{\oplus}/d_r)^2$. For HD 209458b, $d_r$ is approximately 0.047 AU. 

HD 189733b orbits a star of type K1.5V. The closest analogue that could be found for this was the spectrum of epsilon Eridani, a K2V star, for wavelengths above 115 nm (\citealt{Youngblood2016}), and the solar flux below 115 nm. Since HD 189733b orbits at a distance of 0.031 AU, the spectrum is corrected to account for this in the same way as described above for HD 209458b. 

Both these spectra are shown in Figure \ref{fig:Flux}.

\subsection{Chemical Network} \label{sec:chemnet}
\textsc{Levi} is able to use an external chemical network to compute the reactions in the atmosphere, and the rate that at which they occur. In this work, the chemical network being used is a subset of the \textsc{Stand2018} Atmospheric Chemical Network first developed in \cite{R2016}. The version being used is a \ce{H}/\ce{C}/\ce{N}/\ce{O}/\ce{He} network, containing approximately 150 different chemical species and 1000 forward reactions. To improve the generality of the code it is possible to selectively disable certain reaction types, allowing the code to be applied to a much wider range of situations using the same network. The reactions that the model uses can be split into three general types: Bimolecular, termolecular and photochemical.

\subsubsection{Bimolecular Reactions} \label{sec:2-body}
The general form of a bimolecular reaction is,
\begin{equation}
A + B \rightarrow C + D,
\label{eq:2body}
\end{equation}
for which the reaction rate is calculated using the generalised Arrhenius equation,
\begin{equation}
k = \alpha \left(\frac{T}{300}\right)^{\beta}\exp{\left(-\frac{\gamma}{T}\right)},
\end{equation}
where $k$ (\si{\metre\cubed\per\second}) is the rate coefficient, $T$ (K) is the temperature and $\alpha$ (\si{\metre\cubed\per\second}), $\beta$ and $\gamma$ (K) are specific constants for each reaction, determined by fitting the Arrhenius equation to experimentally found or theoretically calculated values of $k$. Reactions of this form, and their rate constants, are tabulated in a supplementary file with the label A or RA, and are numbered 194 - 973 and 1040 - 1044.

\subsubsection{Termolecular Reactions} \label{sec:3-body}
Termolecular reactions typically occur as either decomposition reactions,
\begin{equation}
A + M \rightarrow C + D + M,
\label{eq:decompostion}
\end{equation}
or combination reactions,
\begin{equation}
A + B + M \rightarrow C + M,
\label{eq:combination}
\end{equation}
where the M represents any third body that is involved in the reaction but is chemically unaffected as a result, often acting as either an energy source or sink for decomposition or combination reactions, respectively. To calculate the rate coefficients for these reactions, the Lindemann form is used  (\citealt{Lindemann1922}); this form separates the rate constants into high and low pressure limits,
\begin{equation}
\begin{aligned}
k_0 &= \alpha_0 \left(\frac{T}{300}\right)^{\beta_0}\exp{\left(-\frac{\gamma_0}{T}\right)},
\\
k_{\infty} &= \alpha_{\infty} \left(\frac{T}{300}\right)^{\beta_{\infty}}\exp{\left(-\frac{\gamma_{\infty}}{T}\right)},
\end{aligned}
\end{equation}
with $k_0$ (\si{m^{6}.s^{-1}} for combination or \si{m^{3}.s^{-1}} for decomposition) as the low-pressure rate constant and $k_{\infty}$ (\si{m^{3}.s^{-1}} for combination or \si{s^{-1}} for decomposition) as the high-pressure rate constant. These can be combined to form an effective two-body reaction rate,
\begin{equation}
k = \frac{k_0 [M]}{1+\frac{k_0 [M]}{k_{\infty}}} ,
\end{equation}
where $[M]$ (\si{\per\metre\cubed}) is the number density of the third species, where the third species can be any molecule in the atmosphere. Reactions of this form, and their rate constants, are tabulated in a supplementary file with the label B, and are numbered 1 - 193.

\subsubsection{Reverse reactions}
It is necessary to consider that the reactions described in the previous sections will also occur in the opposite direction, i.e., the previous products will react together to form the reactants. The rate at which this happens can be calculated either with the Arrhenius equation, with the constants being determined experimentally as before, or by thermodynamically reversing the reaction (\citealt{Burcat2005}). Due to the lack of experimental data for these reverse reactions, especially at the temperatures being considered, we are using thermodynamic reversal to determine the reaction rate.

The equation used to calculate the reverse reaction rate is,
\begin{equation}
\frac{k_f}{k_r} = K_c \left(\frac{k_b T}{P_0}\right)^{-\Delta \nu},
\end{equation}
in which $k_f$ and $k_r$ are the rate of the forward and reverse reactions, $k_b$ is the Boltzmann constant, $\Delta \nu$ is the number of products minus the number of reactants, $P_0$ is the standard-state pressure (which is 1 bar, under the ideal gas assumption). $K_c$ is an equilibrium constant determined by,
\begin{equation}
K_c = \exp{\left( -\frac{\Delta G^o}{\mathcal{R}T}\right)},
\end{equation}
where $\mathcal{R}$ is the gas constant and $\Delta G^o$ is the standard state gibbs free energy change on reaction, which expressed as enthalpy and entropy changes on reaction is,
\begin{equation}
\Delta G^o = \Delta H^o - T \Delta S^o,
\end{equation}
with $\Delta H^o$ and $\Delta S^o$ as the standard state change in enthalpy and entropy, respectively, upon reaction. The enthalpy and entropy of each species can be calculated by a series in terms of the NASA polynomials,
\begin{equation}
\begin{aligned}
\frac{H^o}{\mathcal{R}T} &= a_1 + \frac{a_2}{2} T + \frac{a_3}{3}T^2 + \frac{a_4}{4}T^3 + \frac{a_5}{5}T^4 + \frac{a_6}{T},
\\
\frac{S^o}{\mathcal{R}} &= a_1 ln T + a_2 T + \frac{a_3}{2}T^2 + \frac{a_4}{3}T^3 + \frac{a_5}{4}T^4 + a_7.
\end{aligned}
\end{equation}

The values for these polynomials were drawn from Burcat \footnote{http://garfield.chem.elte.hu/Burcat/burcat.html} whenever possible, and when not possible, used a modified Benson's method, described in the Appendix of \cite{R2016}.

\subsubsection{Photochemistry} \label{sec:Photochem}

The rate of photo-dissociation for photochemically active molecules is also calculated. The molecular absorption cross-sections, $\sigma_a$ (\si{\metre\squared}), for these molecules were drawn primarily from PhIDRates \footnote{http://phidrates.space.swri.edu/}, apart from \ce{C4H2} and \ce{C4H4}, which were not in this database. These two molecules used the cross-sections from the MPI-Mainz-UV-VIS Spectral Atlas of Gaseous Molecules \footnote{www.uv-vis-spectral-atlas-mainz.org}. These cross-sections are often temperature dependant, but there is a significant lack of data at high temperatures, and so often the cross-sections we use are measured at room temperature. There has been some work done in identifying the UV cross-section for \ce{CO2} at high temperatures (\citealt{Venot2013,Venot2018}), in which several orders of magnitude difference were found in the absorption cross-section, resulting in significant differences to the computed chemical composition. This emphasises the need for more high temperature cross-sections to accurately compute the photo-chemistry occuring in hot Jupiter atmospheres. The cross-sections were divided into 3000 bins, each 0.1 nm wide, to cover the approximately 300 nm range over which most cross-sections are defined. The photo-dissociation rate is then,
\begin{equation}
k = t_f \int \sigma_a(\lambda) F(\lambda, z) d\lambda 
\end{equation}
with $F(\lambda,z)$ as the Actinic Flux as described in \ref{sec:Flux}, and $t_f$ a dimensionless factor accounting for the fraction of time the atmosphere being modelled spends receiving stellar irradiation. This is 0 or 1 for a tidally locked planet's night and day sides respectively, and on average 1/2 for a planet with a day-night cycle. Photo-dissociation reactions are not reversed, since, while it is possible for reactions of this form to occur, they occur too slowly to have any impact on the atmosphere. Photo-dissociation reactions and their rate constants are tabulated in a supplementary file with the label V, numbers 974 - 1039.

\subsection{Numerical Method}
Here we describe how the governing equations presented in the previous section are discretised into a form that can be solved numerically.

\subsubsection{Discretisation}
Equation \ref{eq:continuity} can be rewritten in a discrete form that allows it to be solved numerically,
\begin{equation}
\frac{\Delta n_{i,j}}{\Delta t} = \mathcal{P}_{i,j} - \mathcal{L}_{i,j} - \frac{ \Phi_{i,j+1/2} - \Phi_{i,j-1/2}}{\Delta z},
\label{eq:discretecontinuity}
\end{equation}
where, as previously, the subscript $i$ refers to the $i^{th}$ species. The subscript $j$ represents the $j^{th}$ layer in the atmosphere, and takes values from $j$ = 1,\dots,$N_z$, with $N_z$ being the total number of atmosphere layers. The diffusion terms have subscripts $j+1/2$ and $j-1/2$ and thus are defined to be at the upper and lower boundary of each layer. Here it has been assumed that only adjacent layers will contribute to the incoming fluxes, as a first-order treatment of diffusion is being used (\citealt{Hu2012}). The layer thickness is given by $\Delta z$. Equation \ref{eq:diffusion} can be similarly discretised to
\begin{multline}
\Phi_{i,j\pm1/2} = \mp(K_{zz,j\pm1/2}+D_{i,j\pm1/2})\ n_{t,j\pm1/2}\ \frac{ X_{i,j\pm1} -X_{i,j}}{\Delta z}\\ + D_{i,j\pm1/2}\ n_{i,j\pm1/2}\left[\frac{(m_{t,j\pm1/2} - m_i)g}{k_b T_{j\pm1/2}}
\mp \frac{\alpha_{T,i}}{T_{j\pm1/2}}\frac{T_{j\pm1} - T_j}{\Delta z}\right],
\label{eq:discretediffusion}
\end{multline}
in which any $Y_{j\pm1/2}$ has been approximated to $(Y_{j\pm1} + Y_{j})/2$. $m_{t,j}$ is the mean mass of atmosphere layer j, and $m_i$ is the mass of species i, $g$ is the gravitational acceleration, and $k_b$ is the Boltzmann constant. Now, it is possible to combine Equation \ref{eq:discretecontinuity} and \ref{eq:discretediffusion} to form a set of ordinary differential equations with time as the independent variable,
\begin{multline}
\frac{\Delta n_{i,j}}{\Delta t} =  \mathcal{P}_{i,j} - \mathcal{L}_{i,j} + \frac{1}{\Delta z_{j-1/2}}[(k^-_{i,j+1/2}\frac{n_{t,j+1/2}}{n_{t,j+1}})\ n_{i,j+1} -\\ (k^+_{i,j+1/2} \frac{n_{t,j+1/2}}{n_{t,j}} + k^-_{i,j-1/2}\frac{n_{t,j-1/2}}{n_{t,j}})\ n_{i,j} + (k^+_{i,j-1/2}\frac{n_{t,j-1/2}}{n_{t,j-1}})\ n_{i,j-1}],
\label{eq:discretion}
\end{multline}
in which
\begin{multline}
k^\pm_{j+1/2} = \frac{K_{zz,j+1/2} + D_{i,j+1/2}}{\Delta z_{j}} \pm \\ \frac{D_{i,j+1/2}}{2\Delta z_j}\left[\frac{(m_{t,j}-m_i)g\ \Delta z_j}{k_b T_{j+1/2}} - \frac{\alpha_{T,i}}{T_{j+1/2}}(T_{j+1}-T_j)\right]
\end{multline}

We define the atmosphere layers to be spaced equally in log pressure, and thus to find the corresponding altitude hydrostatic equilibrium is assumed,
\begin{equation}
P(z) = P_0\ e^{\frac{-m(z)gz}{k_b T(z)}},
\end{equation}
with $m$ as the mean mass of the molecules in an atmosphere layer, found by $\sum\limits_i m_i n_i$, where $m_i$ is the mass of a molecule of species $i$. $g$ is the surface gravity, defined for gas giants as the gravity at 1 bar, which is set to 10 \si{\metre\per\second\squared} for an isothermal profile, 21.4 \si{\metre\per\second\squared} for HD 189733b and 9.36 \si{\metre\per\second\squared} for HD 209458b in accordance with \cite{Moses2011}. The size of the atmospheric layers are recalculated at every time step to ensure the atmosphere stays in hydrostatic equilibrium, since the mass of each layer varies as the system evolves. 

\subsubsection{Solver}
\textsc{Levi} uses a semi-implicit second-order Rosenbrock solver to find a steady-state solution to Equation \ref{eq:discretion}, (\citealt{Rosenbrock1963}). \cite{Verwer1998} show that the second-order Rosenbrock solver is stable over large range of step-sizes. This property of the solver is ideal for chemical kinetics, where step-sizes can range over many orders of magnitude due to the possibility of chemical reactions which have sub-microsecond timescales and diffusion which can occur over timescales of hundreds of years. The general differential equation is,
\begin{equation}
\frac{d n}{d t} = f(n),
\label{eq:ODE}
\end{equation}
where $n$ is the dependent variable, $t$ is the independent variable and $f(n)$ is some general function of $n$. The Rosenbrock solver provides a solution to Equation \ref{eq:ODE} in the form,
\begin{equation}
\begin{aligned}
n_{k+1} &= n_k + 1.5 \Delta t g_1 + 0.5 \Delta t g_2,
\\
(\textbf{I} - \gamma \Delta t \textbf{J})g_1 &= f(n_k),
\\
(\textbf{I} - \gamma \Delta t \textbf{J})g_2 &= f(n_k + \Delta t g_1) - 2 g_1,
\end{aligned}
\label{eq:Solve}
\end{equation}
where $\Delta t$ is the size of the time-step, $\gamma$ is a scaler constant, suggested by \cite{Verwer1998} to be $\gamma=1+1/\sqrt[]{2}$ for stability, the subscript $k$ denotes the $k^{th}$ step in the solver, $\textbf{I}$ is the identity matrix, and
\begin{equation}
\textbf{J} \equiv \frac{\partial f}{\partial n}|_{n_k},
\label{eq:jac}
\end{equation}
is the Jacobian matrix evaluated at $n_k$. 

To solve Equation \ref{eq:Solve} it is necessary to invert the Jacobian matrix, which is one of the most expensive parts of the calculations, and so we focused on making this as efficient as possible. The Jacobian takes the form of a tridiagonal block matrix of total size $N_i N_z$, the product of the total number of species and the total number of atmosphere layers. The construction of this matrix is presented in equation \ref{eq:jacobian}. The diagonal blocks, B, of the Jacobian are made up of square matrices of size $N_i$ that describe the interactions between species in the same atmosphere layer. The off-diagonal blocks, A and C, also matrices of size $N_i$, represent the effects of adjacent atmosphere layers via diffusion. The subscripts for the A, B and C matrices show which atmosphere layer they represent.
\begin{equation}
J = 
\begin{bmatrix}
    B_{1}  & C_{1} & 0 & 0&\dots &0&0&0& 0 \\
    A_{2} & B_{2} & C_{2} & 0 & \dots &0&0&0& 0 \\
    0 & A_{3} & B_{3} & C_{3} & \dots &0&0&0& 0 \\
    0 & 0 & A_{4} & B_{4} & \dots &0&0&0& 0 \\
    \vdots & \vdots & \vdots & \vdots & \ddots &\vdots & \vdots & \vdots & \vdots \\
    0   & 0 & 0 &  0 & \dots & B_{N_z -3} & C_{N_z - 3} & 0 & 0 \\
    0   & 0 & 0 &  0 & \dots & A_{N_z -2} & B_{N_z - 2} & C_{N_z -2} & 0 \\
    0   & 0 & 0 &  0 & \dots & 0 & A_{N_z -1} & B_{N_z - 1} & C_{N_z -1} \\
    0   & 0 & 0 &  0 & \dots & 0 & 0 & A_{N_z} & B_{N_z} \\
\end{bmatrix}
\label{eq:jacobian}
\end{equation}

By comparison of equations \ref{eq:prodloss}, \ref{eq:discretion}, \ref{eq:ODE}, and \ref{eq:jac}, it can be seen that
\begin{equation}
\begin{aligned}
A &= \frac{k^-_{i,j+1/2}\ n_{t,j+1/2}}{dz_{j-1/2}\ n_{t,j+1}},
\\
C &= \frac{k^+_{i,j-1/2}\ n_{t,j-1/2}}{dz_{j-1/2}\ n_{t,j-1}}, 
\\
B &= -\left(\frac{k^+_{i,j+1/2}\ n_{t,j+1/2}}{dz_{j-1/2}\ n_{t,j}}+\frac{k^-_{i,j-1/2}\ n_{t,j-1/2}}{dz_{j-1/2}\ n_{t,j}}\right) 
\\
&+\frac{\partial}{\partial n_{j}} (\sum (k_2\ n_{i''}\ n_{i'''} - k_1\ n_i\ n_{i'})).
\end{aligned}
\end{equation}
As a result, both A and C are purely diagonal matrices, while B is a full matrix with each position describing how one species reacts with another in the same layer. 

To invert this Jacobian, we use the method of inverting a tridiagonal matrix described in \cite{Hubeny2014}, with a generalisation to block tridiagonal by treating each matrix element in the inversion method as a full matrix. As a result, only the three $N_i$ size matrices are ever kept in the memory, which is much less expensive than trying to invert the complete $N_i N_z$ matrix.

\subsubsection{Convergence}

There are several conditions that must be satisfied before the run is considered to be complete and have reached steady-state. These are that both the relative change, $\Delta n = \frac{|n_{i,j,k+1} - n_{i,j,k}|}{n_{i,j,k}}$, and the rate of relative change, $\frac{\Delta n}{\Delta t}$ of mixing ratios in the atmosphere, are sufficiently small that running the code further would produce no significant change.

To ensure that each step is correctly approaching steady-state, several conditions are checked after every time-step. If the conditions are passed then the solver continues, otherwise, the latest solution is discarded and the step-size is reduced, before re-running the solver. The conditions are that every species has a positive number density and that the truncation error, $\epsilon$ is sufficiently small.

The truncation error is found by calculating the difference between second-order solution and the first-order solution, $n^1_{k+1} = n_k + \Delta t g_1$, such that,
\begin{equation}
\epsilon = |n_{k+1} - n^1_{k+1}|.
\end{equation}

Since the timescales being modelled can vary over many orders of magnitude, having a time-step that is self-adjusting is vital to allowing the code to have a reasonable runtime. The truncation error can thus be used as a cheap local error estimation to control the step size,
\begin{equation}
\Delta t_{k+1} \sim \Delta t_{k} (1/\epsilon)^{0.5}.
\end{equation}

\subsection{Initial and Boundary Conditions}

\subsubsection{Initial Conditions}
\label{sec:Initial}
In the present work we focus on hydrogen-rich gas giant atmospheres. Our baseline model uses solar elemental abundances, from \cite{Asplund2009}. In Section \ref{Section4} we also explore the effects of variation in the \ce{C}/\ce{O} and \ce{N}/\ce{O} ratios. In \cite{Asplund2009} the solar abundances are given as fractions of the number of hydrogen molecules, while for \textsc{Levi} the elemental abundances have been transformed to the fraction of the total number of molecules (i.e., the mixing ratio), enabling easier description of non-hydrogen dominated atmospheres. The ratio of elemental atoms to the total number of molecules therefore starts as: $X_{\ce{H}} = 1.707$, $X_{\ce{He}} = 0.145$, $X_{\ce{C}} = 4.594\times 10^{-4}$, $X_{\ce{N}} =1.154\times 10^{-4} $ and $X_{\ce{O}} = 8.359\times 10^{-4}$. When calculating solar abundances all \ce{H} is assumed to be in the form of \ce{H2}, thus leading to a $X_{\ce{H}}$ that is greater than one. This gives a \ce{C}/\ce{O} ratio of 0.55 and a \ce{N}/\ce{O} ratio of 0.138.

When choosing the initial conditions, conservation of mass is ensured by summing over the product of the species present ($n_{i}$) and the number of atoms of $x$ contained by $n_i$ ($A_{i,x}$),

\begin{equation}
\sum\limits_{i} A_{i,x} n_i = X_x .
\label{eq:Consveration}
\end{equation}

The initial chemistry of the atmosphere is chosen as a set of species that contain all the elements in the atmosphere, then by using Equation \ref{eq:Consveration} for each element to solve for the species' initial abundance and setting the initial abundance of all other species in the network to zero. To simplify solving Equation \ref{eq:Consveration}, we picked the initial species set to be small molecules, so that dividing the elements between them would be straightforward.

\begin{figure*}
    \centering 
    \includegraphics{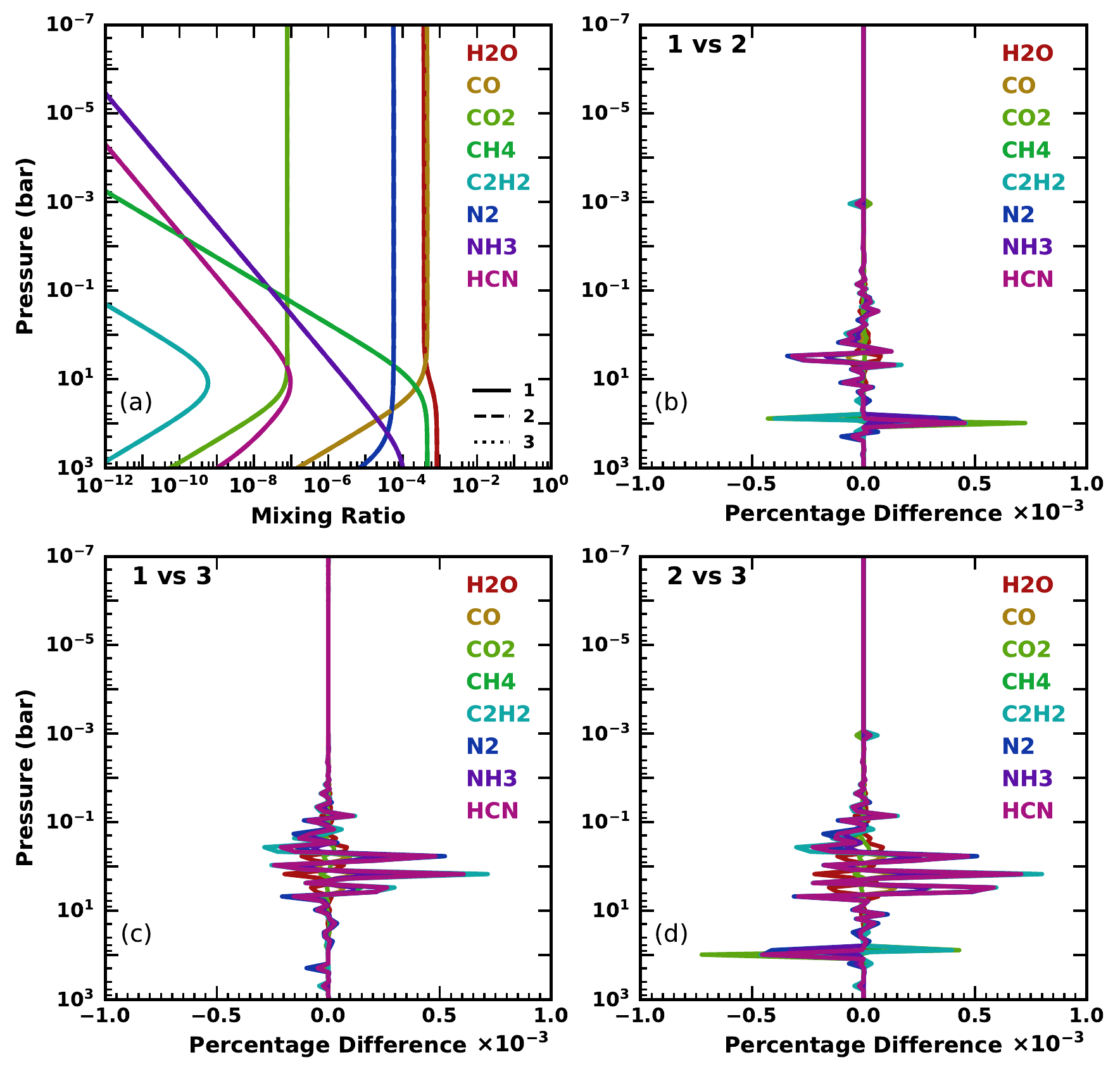}
    \caption[Initital Condition Comparison]{An investigation of the effect of initial conditions on atmospheric chemistry. The sets of initial species are: \ce{H2}, \ce{CO}, \ce{OH} and \ce{NH3} (Solid lines, labeled 1; \ce{H2}, \ce{CH4}, \ce{H2O} and \ce{N2} (Dashed lines, labeled 2; and elements in their atomic form (Dotted lines, labeled 3. In this example a 1500K Isotherm was used, with no diffusion or photochemistry. Plot a) shows all three starting conditions, while b), c) and d) each show the percentage difference between two of the starting options. }
    \label{fig:Initial}
\end{figure*}

With this approach, there are still a range of viable starting conditions for the atmospheric mixing ratio and it is important to demonstrate that model convergence is not affected by this decision. A comparison of the steady state solution emerging from a choice of three initial atmospheric species sets is shown in the first plot of Figure \ref{fig:Initial}, with the initial species sets being: (1) \ce{H2}, \ce{CO}, \ce{OH} and \ce{NH3}; (2) \ce{H2}, \ce{CH4}, \ce{H2O} and \ce{N2}; and (3) atomic \ce{H}, \ce{C}, \ce{O}, and \ce{N}. Having the initial species set being elements in their atomic form provides a starting condition of extreme disequilibrium, compared to starting with molecules that are expected to dominate the steady-state atmosphere, as in the other two starting conditions. There is no noticeable difference between the three sets of mixing ratios, confirming that the steady-state solution is independent of the choice of atmospheric species set to initialise the calculation with. The time taken to reach this steady state, however, is significantly different between the three sets, with set (1) taking four and three times as long as sets (2) and (3) respectively, in this scenario. Plots b), c) and d) of Figure \ref{fig:Initial} show the accuracy to which each of the three different initial conditions are the same. In all cases there was less than a 10 ppm difference between any two starting sets of molecules, thus suggesting that the solutions these initial sets produce are functionally identical.  

\subsubsection{Boundary Conditions}
\label{sec:Boundary}
It is necessary to specify boundary conditions at the minimum and maximum pressures chosen for the simulation. There are a variety of choices available, depending both on the type of planet and the range of pressures being explored. 

For terrestrial exoplanets, the lower boundary conditions usually model any surface-atmosphere interaction, so typical boundary conditions can include permanently assigning a species an abundance at the surface to model a large reservoir, or a flux to describe surface emission and deposition. For the upper atmosphere a flux due to atmospheric escape is common (\citealt{Hu2012}). 

Gas-giant exoplanets have no solid surface at the lower boundary, thus the boundary conditions are either chemical equilibrium or zero flux. The upper boundary condition could be atmospheric escape or, if the upper boundary is deep enough in the atmosphere, zero flux. In general, atmospheric escape does not significantly affect the mass of the atmosphere (\citealt{Linsky2010,Murrayclay2009}). For a boundary in chemical equilibrium, the assumption is that layers deeper in the atmosphere have a chemical timescale much shorter than the dynamical timescale, so the boundary would be in chemical equilibrium. A zero flux, or closed boundary, assumes that the deeper layers are chemically identical to the bottom layer in the calculation, such that there will be no chemical consequence to flow of molecules across the layer boundaries. 

\begin{figure*}
    \centering
	\includegraphics{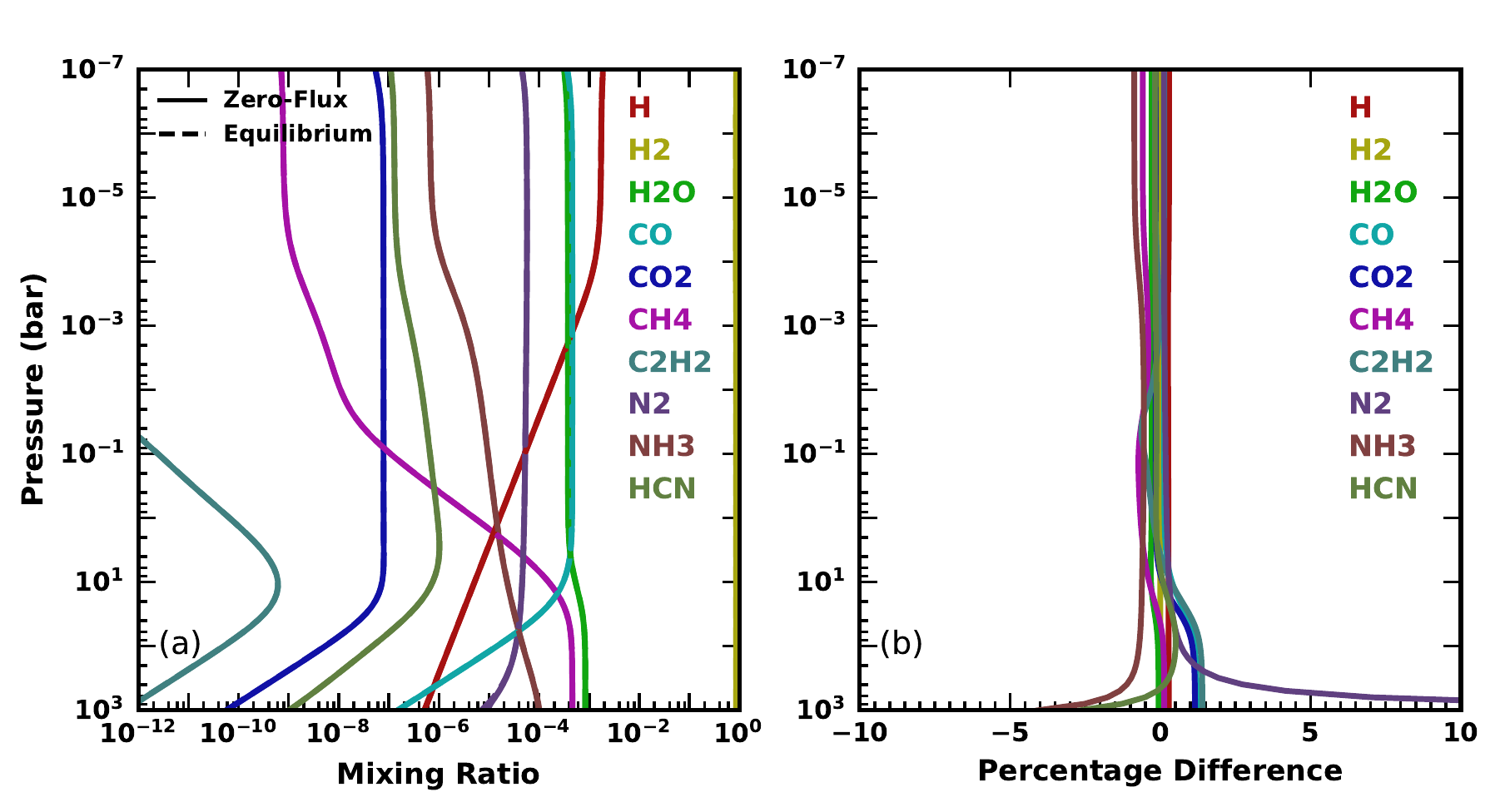}
    \caption[Boundary Condition Comparison]{The difference in steady-state atmospheric chemistry based on the lower boundary condition. In this example a $1500\ \mathrm{K}$ Isotherm was used, with $1\times10^{6}$ \si{\metre\per\second\squared} $K_{zz}$ and no photochemistry. Plot (a) shows a comparison of the atmospheric chemistry between the two boundary conditions, while (b) shows the percentage difference in the abundances of a zero-flux boundary condition compared to a equilibrium boundary condition.}
     \label{fig:Boundary}
\end{figure*}

The effect of setting different lower boundary conditions are explored in the left plot in Figure \ref{fig:Boundary}, which shows the difference between equilibrium (Dashed lines) and zero flux (Solid lines) boundary conditions. It can be seen that there is a slight difference in the mixing ratios as a result of these boundary condition changes, especially to \ce{N2}. To better quantify this change, the right plot in Figure \ref{fig:Boundary} shows the percentage difference between zero flux and equilibrium boundary conditions. Throughout most of the atmosphere the difference in abundance is less than one percent, however closer to the lower boundary it can increase up to 10\%. Thus, it is clear that the boundary conditions can have a much more significant impact upon atmospheric abundances than the initial conditions, although even a 10\% change is negligible when considering the precision with which the abundance of these molecules can be detected in exoplanet atmospheres. Since it is possible for diffusion to drive even the deepest layers slightly away from equilibrium, zero flux boundary conditions were chosen for this work, to prevent a disparity between the boundary layer and the layer above. As a result, zero flux conditions were also chosen for the upper boundary, to ensure conservation of mass in the atmosphere.

\section{Validation and Testing} \label{Section3}

In this section we apply our model to the atmospheres of several hot Jupiters. We first compare our models of HD 209458b to that of prevoius codes (\citealt{Moses2011} and \citealt{Venot2012}) for validation. We then investigate the effects of disequilibrium chemistry on the atmospheric chemistry of hot Jupiters and discuss why these effects occur. We also show the fastest reaction pathway the model takes for several important net reactions in the atmosphere. These pathways are found by use of a code that traces the reactions that lead from the products to the reactants, and identifying the route that has the fastest overall reaction rate. When these pathways are shown, the bracketed numbers adjacent to the reaction correspond to its position in the chemical network, as described in \ref{sec:chemnet}.

\subsection{Validation}
\label{sec:val}

\begin{figure*}
    \centering
    \includegraphics{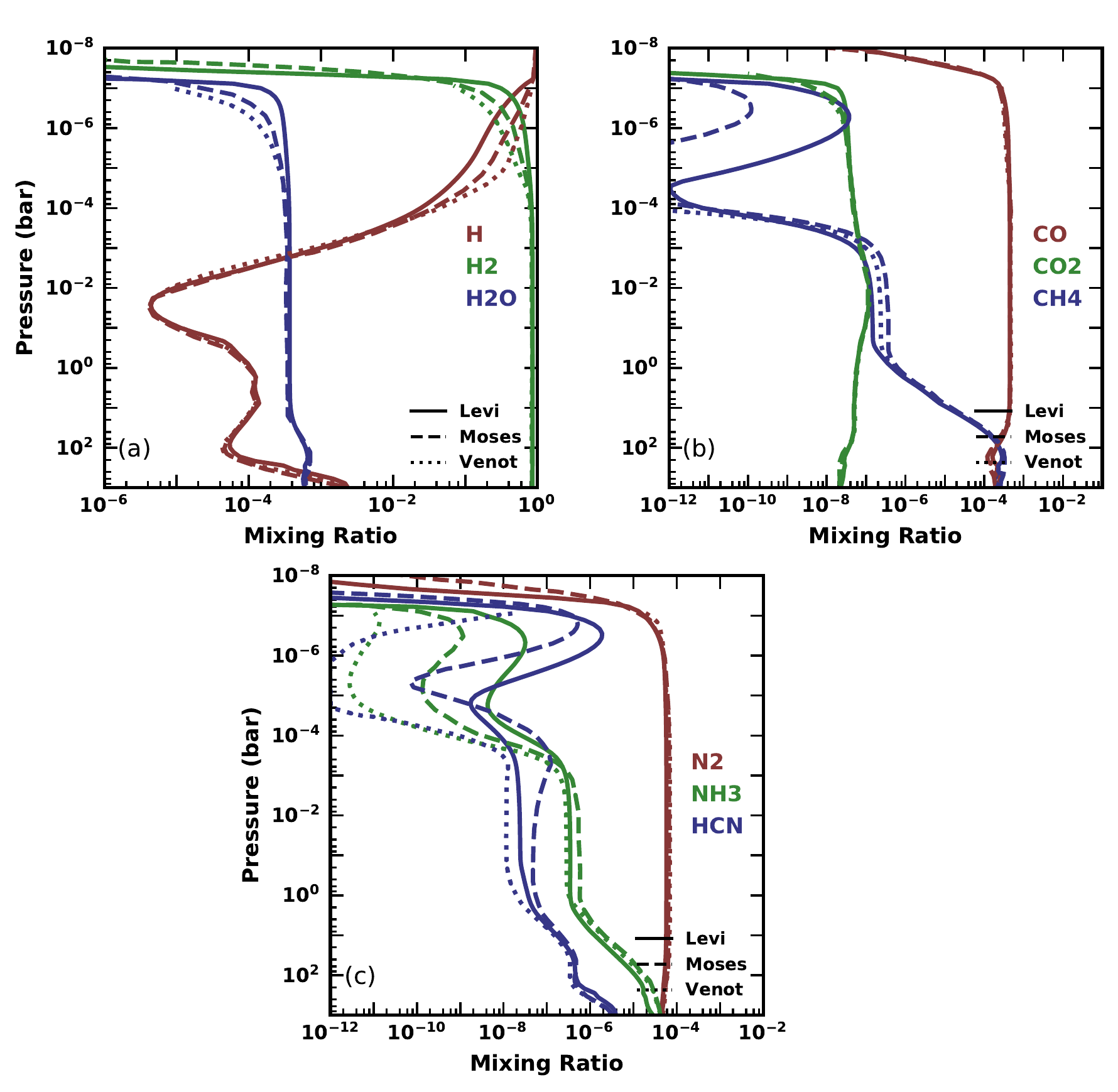}
    \caption[Benchmarking against HD 290458b]{A comparison of models of the atmosphere of HD 209458b for a selection of major molecules in the atmosphere. The solid line is the model described in this paper, the dashed line is the model created by \cite{Moses2011} and the dotted line is the code described in \cite{Venot2012}}
    \label{fig:AllComp}
\end{figure*}

We chose HD 209458b as one of the best candidates for benchmarking \textsc{Levi}, due to both the well characterised spectral observations of its atmosphere and the existence of several previous chemical models that have predicted atmospheric compositions for this planet (\citealt{Moses2011}, \citealt{Venot2012}, \citealt{Agundez2014b}, \citealt{R2016}, \citealt{Tsai2017}). The results produced by this code are compared to the disequilibrium models of HD 209458b of both \cite{Moses2011} and \cite{Venot2012} (Figure \ref{fig:AllComp}). Like \textsc{Levi}, both other models do not self-consistently calculate the P-T profile, but rather use a fixed input. The species chosen for the comparison are \ce{H2}, \ce{H} and \ce{H2O} (Figure \ref{fig:AllComp}a), \ce{CO}, \ce{CH4} and \ce{CO2} (Figure \ref{fig:AllComp}b), and \ce{N2}, \ce{NH3} and \ce{HCN} (Figure \ref{fig:AllComp}c). These species are expected to be some of the most abundant (e.g. \citealt{Swain2009,Macdonald2017}), and several have strong spectral signatures.

Looking at the comparison plots in Figure \ref{fig:AllComp}, it can be seen that for most species there is no significant difference between \textsc{Levi} and \cite{Venot2012}, or between \textsc{Levi} and \cite{Moses2011}. In the deep atmosphere there are some slight differences in the abundances of many molecules. These differences are not significant, and are likely a consequence of slightly differing P-T profiles, thermodynamic coefficients, and choices of initial elemental ratios. For example, \cite{Moses2011} and \cite{Venot2012} model a lower-oxygen atmosphere, because they have considered 20\% of the oxygen abundance to be sequestered in silicates.

One of the largest differences between the models is for \ce{NH3}, where there is an half an order-of-magnitude lower quenching abundance of \ce{NH3} for \textsc{Levi} compared to \cite{Moses2011}, while \textsc{Levi} predicts nearly a hundred times more \ce{NH3} at $10^{-6}$ bar. Compared to \cite{Venot2012}, \textsc{Levi} predicts a nearly identical quenching abundance of \ce{NH3}, and almost ten thousand times more \ce{NH3} at $10^{-6}$ bar. These differences are due to a high degree of uncertainty in the rate of the pathway that converts \ce{NH3} to \ce{N2}:

\begin{equation}
\begin{aligned}
\ce{H2 + M &-> H + H + M} \ (R5) \\
2(\ce{H + NH3 &-> NH2 + H2})  \ (R241) \\
2(\ce{H + NH2 &-> HN + H2})  \ (R220) \\
\ce{2HN &-> N2 + 2H}  \ (R383) \\
\mathbf{Net:}\ \ce{2NH3 &-> N2 + 3H2},
\end{aligned}
\end{equation}

The observed difference in the quenching location of \ce{CH4} arises for similar reasons. In the network used by \textsc{Levi}, there are several pathways that convert \ce{CH4} into \ce{CO}, one of which has a highly uncertain reaction rate (\citealt{R2016}), and can explain the nearly 1 dex and 0.5 dex higher quenching abundance of \ce{CH4} seen in \cite{Moses2011} and \cite{Venot2012}, respectively, than in \textsc{Levi}.
The large differences in abundance seen in \ce{HCN} between the three models throughout the atmosphere are due to high degrees of uncertainty in the rate constants of many of the reactions that include \ce{HCN}.

To conclude, \textsc{Levi} can reproduce the composition of HD 209458b predicted by several other chemical models at an acceptable degree of accuracy. While there are still some significant differences in abundances of several highly abundant species in exoplanet atmospheres, these are the result of uncertainties in rate constants and absorption cross-sections, which differ between the models. Only new experiments and calculations of the rate constants and absorption cross-sections will help resolve these differences.

\subsection{Equilibrium vs. Diffusion vs. Photochemistry} \label{sec:evkvp}

\begin{figure*}
    \centering
    \includegraphics{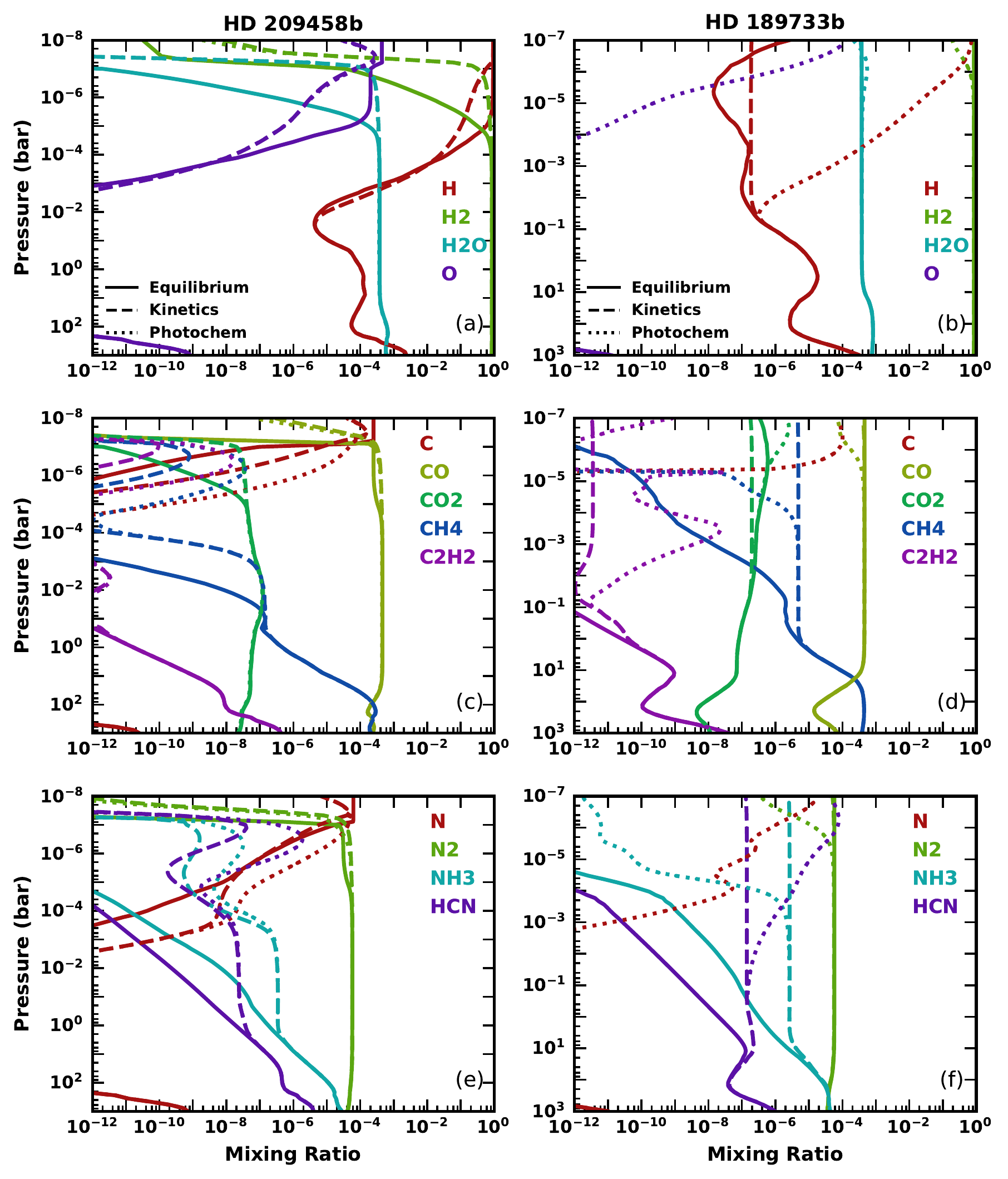}
    \caption[Chemistry and Transport Comparisons]{Abundance profiles for several species on HD 290458b (left) and HD 189733b (right) for purely equilibrium chemistry (solid lines), diffusion (dashed line) and both diffusion and photochemistry (dotted line). The P-T and the $K_{zz}$ profiles for these models are shown in Figures \ref{fig:PT} and \ref{fig:Kzzprof}. }
    \label{figs:evskvsp}
\end{figure*}

In this section the effects of the three main chemical processes that occur in atmospheres are compared and contrasted. Thermochemical equilibrium, diffusion in the form of molecular, thermal and eddy-diffusion and photochemistry can all combine to play an important role in determining the distribution of observable species in exoplanet atmospheres, and thus should be considered when investigating the chemistry of these exoplanets. Chemical equilibrium is expected to dominate in the lower atmosphere, up to a few bars, due to the high temperatures and pressure. Diffusion effects then take over for some species between approximately $1$-$10^{-4}$ bars, and photochemistry dominates above this. At higher altitudes, in the thermosphere, it is sufficiently hot that most species thermally dissociate down to their base elements. We apply our code to the hot Jupiters HD 209458b and HD 189733b and discuss how and why the equilibrium and disequilibrium models diverge. We run three model scenarios, with the results presented in Figure \ref{figs:evskvsp}: no-diffusion, no-photochemistry (solid lines); with diffusion, but no-photochemistry (dashed lines); with diffusion and photochemistry (dotted lines). Results similar to these have also been produced by previous 1-D chemical codes (e.g. \citealt{Zahnle2009,Moses2011,Hu2012,Venot2012,R2016,Tsai2017}).

\subsubsection{Equilibrium}
In the deepest regions of the atmospheres of the planets being investigated, the high temperature and pressure may produce an atmosphere in chemical equilibrium. The relative abundances of species at chemical equilibrium depend on the P-T profile. For the investigated planets, at solar composition, we find that \ce{CH4} and \ce{H2O} dominate over \ce{CO}, and \ce{NH3} approaches the abundance of \ce{N2}, while higher in the atmosphere this trend is reversed (\citealt{Heng2016}). When solving for equilibrium analytically, this same trend is found. In the observable region of the atmosphere, approximately $1$ - $10^{-4}$ bar for emission spectrography, \ce{CO} is the dominant carbon species, with most of the remaining oxygen ending up in \ce{H2O}, and \ce{N2} is the dominant nitrogen species.

On HD 209458b, the thermosphere begins around $10^{-7}$ bar, causing most species to dissociate down to their base atomic forms. The thermosphere of HD 189733b does not begin until a higher altitude, so no thermal dissociation can be seen for this planet in Figure \ref{figs:evskvsp}. Notable for comparison later on are the abundances of \ce{HCN} and \ce{NH3} in the observable region of the atmosphere; in this equilibrium scenario they are very minor constituents, far below detectable limits (\ce{HCN} or \ce{NH3} abundances at least 1\% of the \ce{H2O} abundance) for solar composition (\citealt{Macdonald2017}). 

\subsubsection{Diffusion}
The inclusion of diffuion, the dashed lines in Figure \ref{figs:evskvsp}, leads to some significant deviations from chemical equilibrium. At very high pressures, the temperature is high enough that the chemical time-scale of the reactions is much shorter than the dynamical time-scale, and therefore the atmosphere at these levels is still in chemical equilibrium, as predicted. However, as the pressure and temperature decrease, and reaction rates drop due to the decreased rates of molecule-molecule interactions, the rate of some reactions drop sufficiently such that the time-scale of dynamic motion is now shorter than that of chemical interaction. This results in quenching at the pressure level where the two time-scales are equal, with the species abundance being effectively frozen in at this point and this chemistry being transported higher into the atmosphere by eddy diffusion. This process leads to the possibility of significantly different abundances of some species in the observable region of the atmosphere, compared to chemical equilibrium. 

Figure \ref{figs:evskvsp} shows a number of species being affected by quenching in observable regions of the atmosphere, including \ce{CH4}, \ce{HCN} and \ce{NH3}. \ce{NH3} is of particular note as quenching increases its abundance by many orders of magnitude, on both HD 209458b and HD 189733b, to potentially detectable levels (\citealt{Macdonald2017}). 

On HD 209458b the high temperatures lead to a shorter reaction timescale compared to the mixing timescale, implying that the condition for chemical equilibrium to be in effect is satisfied until much lower pressures in the atmosphere, compared to HD 189733b where quenching occurs deeper in the atmosphere. HD 189733b has many species at their quenched abundance throughout the upper atmosphere. This is due to the the planet's upper atmosphere not having a thermal inversion below the thermosphere, so the chemical timescale monotonically increases with decreasing pressure. On the other hand, HD 209458b does have a thermal inversion, so the chemical timescale can decrease with decreasing pressure, and so the upper atmosphere may begin to return to chemical equilibrium.

\subsubsection{Photochemistry}
The dotted lines in Figure \ref{figs:evskvsp} show how the inclusion of photochemistry affects the atmosphere. The photolysis of \ce{CO}, \ce{H2O}, \ce{N2}, \ce{NH3} and \ce{CH4} drive most of the photochemical reactions that occur in the upper atmosphere.

The destruction of \ce{H2O} by a UV photon into \ce{OH} and \ce{H}, as deep as 0.1 bar, sets up the reaction that turns \ce{H2} into \ce{H}:

\begin{equation}
\begin{aligned}
\ce{H2O + }\ h\nu \ce{ &-> OH + H} \ (R992) \\
\ce{H2 + OH &-> H2O + H}  \ (R221) \\
\mathbf{Net:}\ \ce{H2 + }\ h\nu \ce{ &-> 2H}.
\end{aligned}
\end{equation}

This reaction provides a large pool of highly reactive hydrogen radicals that can diffuse to both higher and lower altitudes to cause a chain of further reactions, drastically altering the composition of the upper atmosphere.  On HD 209458b the rate of this reaction is low compared to the rate of dissociation of molecules due to the high temperatures in the upper atmosphere, and so little difference can be seen between the case with and without photochemistry. As evidenced in Figure \ref{figs:evskvsp}a and \ref{figs:evskvsp}b, \ce{H2O} is not permanently depleted from the atmosphere until much lower pressures due to it being replenished as fast as it is destroyed. 

The photolysis of \ce{CO} can produce \ce{C} and \ce{O} radicals that can also contribute to the destruction of \ce{H2}. Like \ce{H2O}, \ce{CO} is also being replenished quickly, in this case by a reaction between \ce{CH4} and \ce{H2O}:

\begin{equation}
\begin{aligned}
2(\ce{H2 + M &-> 2H +M}) \ (R5) \\
\ce{H2O + H &-> OH + H2}  \ (R221) \\
\ce{CH4 + H &-> H2 + CH3}  \ (R251) \\
\ce{CH3 + OH &-> CH2O + H2}  \ (R479) \\
\ce{H + CH2O &-> H2 + CHO}  \ (R236) \\
\ce{H + CHO &-> CO + H2} \ (R216) \\
\mathbf{Net:}\ \ce{CH4 + H2O &-> CO + 3H2}.
\end{aligned}
\end{equation}

 However, as Figure \ref{figs:evskvsp}d shows, on HD 189733b the formation of \ce{CO} is not efficient enough to prevent loss of \ce{CO} above $10^{-5}$ bar and so some of the carbon can end up in other species, e.g., \ce{CO2} and \ce{HCN}. On HD 209458b, the temperature, and thus the reaction rates, are high enough to keep replenishing \ce{CO} until pressures as low as $10^{-7}$ bar. 

The \ce{C} radicals from the photolysis of \ce{CO} often end up producing \ce{CH4}:

\begin{equation}
\begin{aligned}
\ce{CO + }\ h\nu \ce{ &-> C +O} \ (R979) \\
\ce{C + H2 &-> CH + H}  \ (R195) \\
\ce{CH + H2 &-> CH2 + H}  \ (R212) \\
\ce{CH2 + H2 &-> CH3 + H}  \ (R235) \\
\ce{CH3 + H2 &-> CH4 + H}  \ (R251) \\
\mathbf{Net:}\ \ce{CO + 4H2 &-> CH4 + O + 4H},
\end{aligned}
\end{equation}

\noindent which leads to the increase of methane seen on HD 209458b. \ce{CH4} is, in turn, susceptible to destruction by \ce{H} radicals, thus leading to an overall decrease in methane if sufficient \ce{H} is produced by the photo-dissociation of \ce{H2O}, as discussed earlier, and as can be seen at $10^{-5}$ bar on HD 189733b. 

\ce{CO2} is not greatly affected by photochemistry since its production is based upon fast reactions between \ce{H2O} and \ce{CO}, both of which are largely constant until high in the atmosphere. In the upper atmosphere of HD 189733b, some of the \ce{CO} is transformed into \ce{CO2}

\ce{C2H2} is an important byproduct of photochemistry. The carbon atoms released from dissociation of \ce{CH4} or \ce{CO} are reprocessed to form \ce{C2H2}. This is the result of a highly efficient pathway:

\begin{equation}
\begin{aligned}
\ce{CO + }\ h\nu \ce{ &-> C +O} \ (R979) \\
\ce{C + H2 &-> CH + H}  \ (R195) \\
\ce{CH + H2 &-> CH2 + H}  \ (R212) \\
\ce{CH4 + }\ h\nu \ce{ &-> CH3 +H} \ (R1019) \\
\ce{CH3 + M &-> CH2 + H + M} \ (R29) \\
\ce{CH2 + CH2 &-> C2H2 + H2} \ (R614) \\
\mathbf{Net:}\ \ce{CO + CH4 + 2H2 &-> C2H2 + 4H},
\end{aligned}
\end{equation}

which leads to many orders of magnitude more \ce{C2H2}, on both HD 209458b and HD 189733b, than otherwise expected if there was no UV flux.

Photo-chemistry can also speed up the reaction converting \ce{NH3} to \ce{HCN},
\begin{equation}
\begin{aligned}
\ce{NH3 + }\ h\nu \ce{ &-> NH2 + H} \ (R1013) \\
\ce{NH2 + }\ h\nu \ce{ &-> HN + H}  \ (R999) \\
\ce{H + NH &-> H + H + N}  \ (R204) \\
\ce{CH4 + M &-> CH3 + H + M} \ (R34) \\
\ce{CH3 + N &-> HCN + H2} \ (R303) \\
2(\ce{H + H + M &-> H2 + M}) \ (R5) \\
\mathbf{Net:}\ \ce{NH3 + CH4 &-> HCN + 3H2},
\end{aligned}
\end{equation}
causing a significant increase in the amount of \ce{HCN}. On HD 209458b this proceeds to the point where \ce{HCN} is the second most abundant carbon and nitrogen bearing molecule in the atmosphere, and on HD 189733b it is the most abundant carbon and nitrogen bearing molecule. This reaction also causes the large decrease of \ce{NH3} seen in the upper atmosphere of HD 189733b, compared to models without photochemistry. Although loss processes exist for \ce{HCN}, the products are normally not very stable, and often form \ce{HCN} upon their destruction. The result is that destruction reactions are inefficient in regulating \ce{HCN}'s abundance, especially with other pathways forming it fast enough to replenish any loss. 

\section{Exploration of the Chemical Parameters} \label{Section4}

\begin{figure}|
    \centering
    \begin{subfigure}[t]{0.48\textwidth}
        \centering
        \includegraphics[height=2.3in]{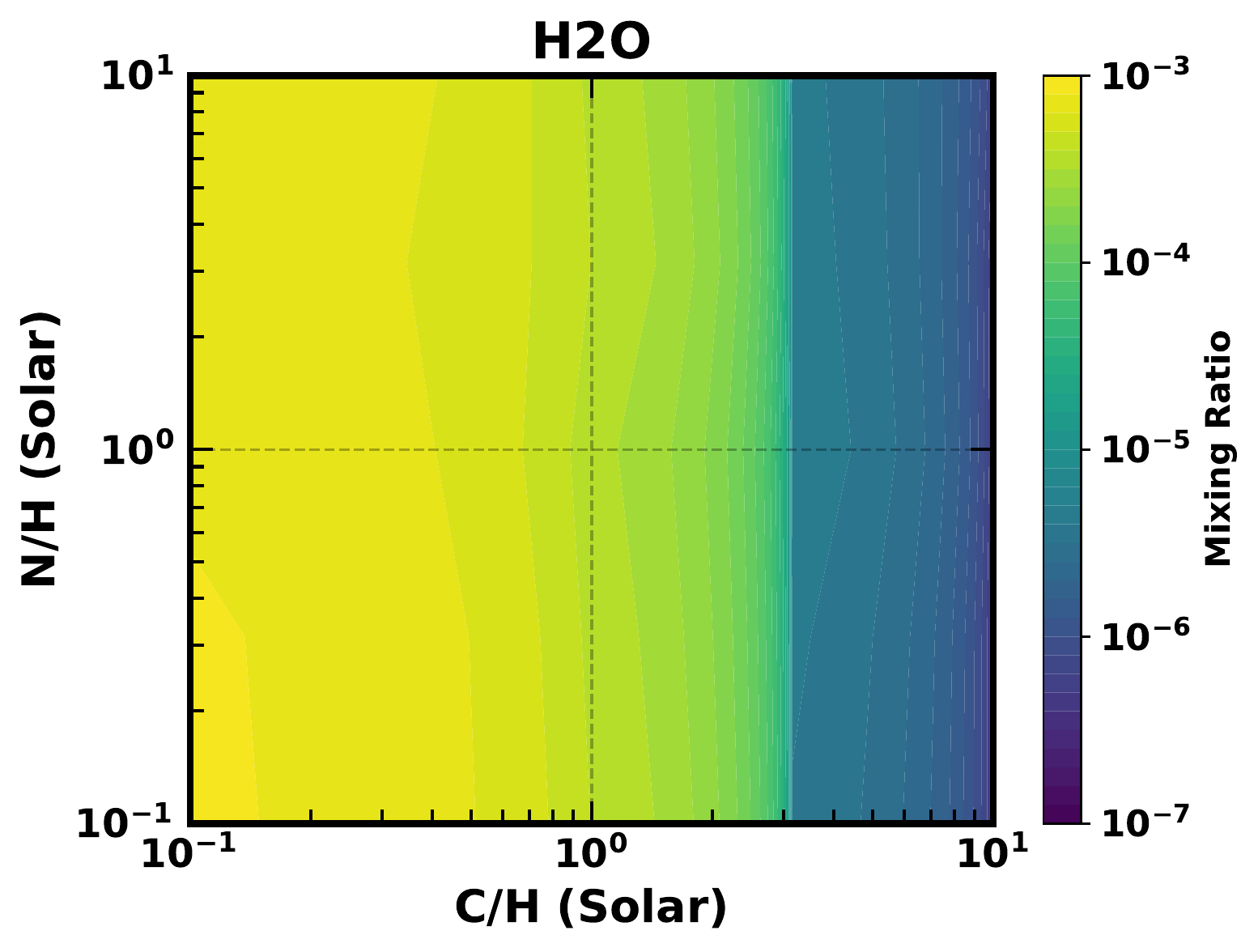}
        \caption{}
        \label{fig:ratioH2O}
    \end{subfigure}
    \begin{subfigure}[t]{0.48\textwidth}
        \centering
        \includegraphics[height=2.3in]{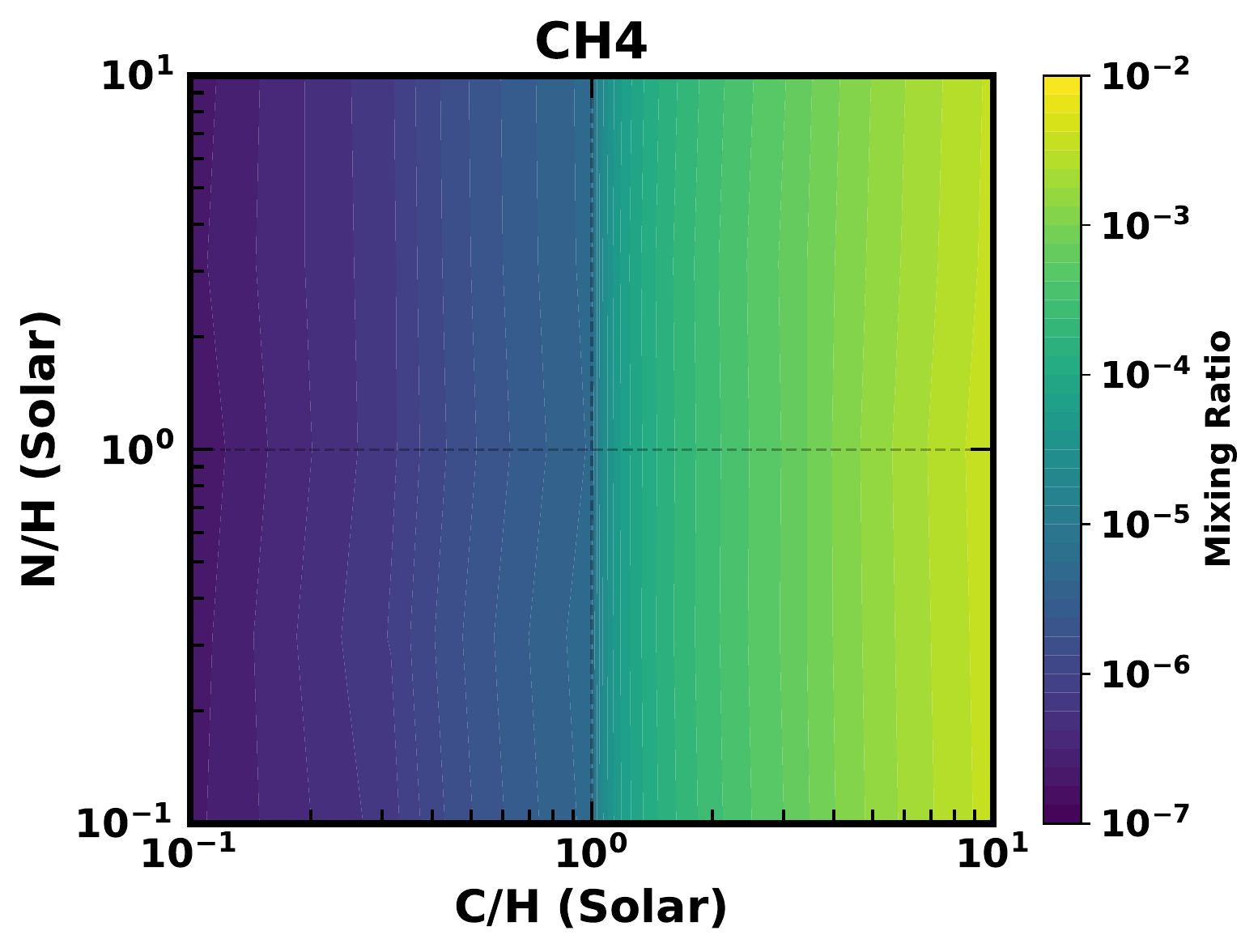}
        \caption{}
        \label{fig:ratioCH4}
    \end{subfigure}
	\begin{subfigure}[t]{0.48\textwidth}
    	\centering
        \includegraphics[height=2.3in]{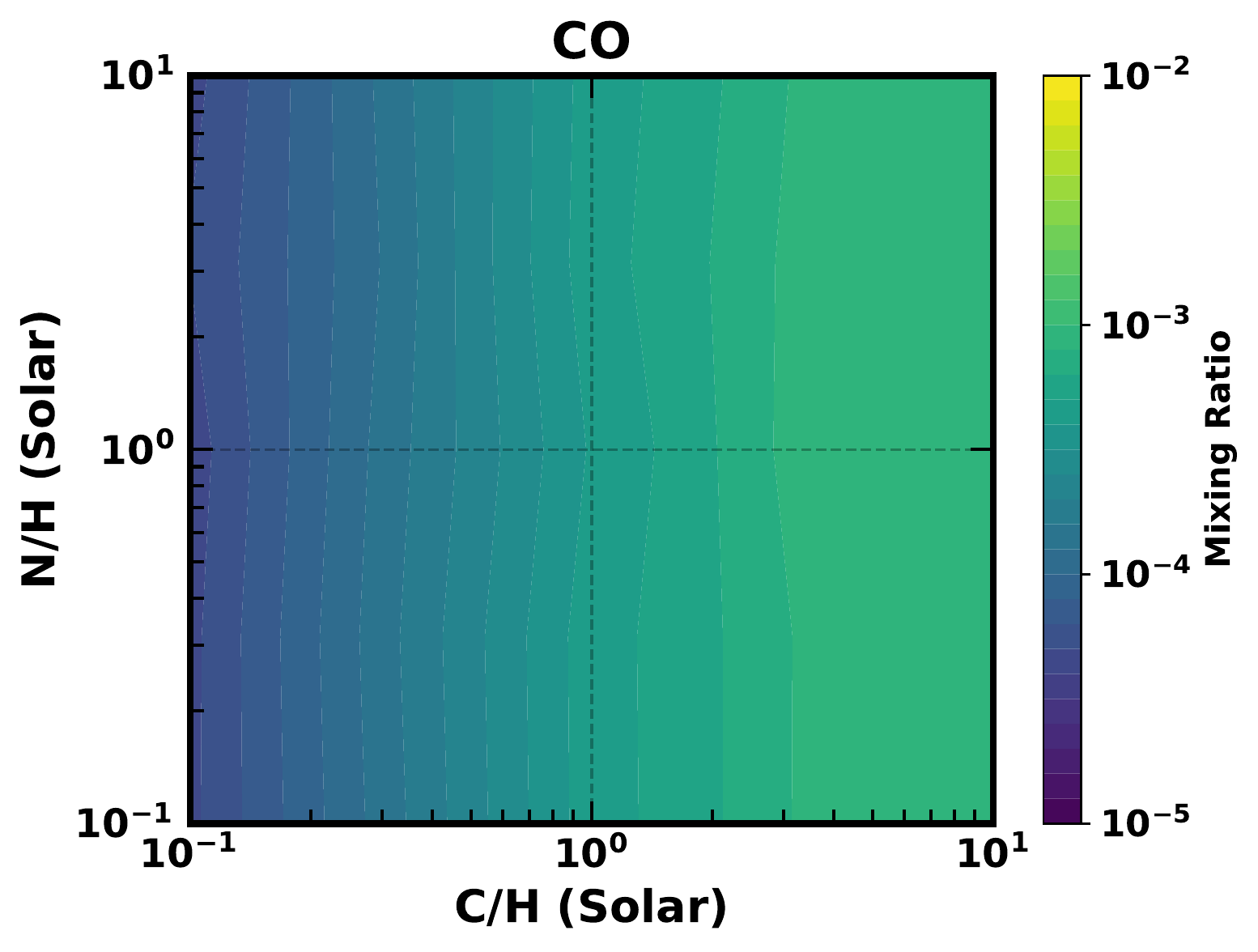}
        \caption{}
        \label{fig:ratioCO}
	\end{subfigure}
    \caption[Variation of carbon and oxygen species with varying C/N/O ratios]{Abundance variations of \ce{H2O} (top), \ce{CH4} (middle) and \ce{CO} (bottom) on HD 189733b at 0.1 bar for the parameter space in which the amount of nitrogen and carbon in the atmosphere varies between 0.1 and 10 times the solar amount. These models use the P-T and diffusion profile from Figure \ref{fig:PT} and \ref{fig:Kzzprof} for HD 189733b.}
    \label{figs:ratios1}
\end{figure}

In this section we perform an initial exploration into the parameter space of the \ce{C}/\ce{O} ratio and \ce{N}/\ce{O} ratio, and consider how it may effect the detectability of certain species in exoplanet atmospheres. Apart from where stated otherwise, these models use the P-T profile and $K_{zz}$ profile described in Section \ref{sec:Atmos}, the initial conditions labelled (a) and the atmospheric composition from Section \ref{sec:Initial} and the zero-flux boundary conditions from Section \ref{sec:Boundary}. Finally, we use a new P-T profile for HD 209458b that does not contain a thermal inversion to investigate the \ce{C}/\ce{O} and \ce{N}/\ce{O} ratios that would best fit with new evidence of multiple species on HD 209458b.

\subsection{The \ce{C}/\ce{O} and \ce{N}/\ce{O} ratio} \label{sec:Ratios}

Throughout our previous models we have used a solar composition for the atmospheres of the planets being investigated (Section \ref{sec:Initial}). However, there is evidence that some hot Jupiters have \ce{C}/\ce{O} ratios substantially different from solar values (\citealt{Madhu2011}, \citealt{Stevenson2014}). A planet's composition depends on many factors, mainly arising from its formation and migration history. Thus knowledge of the atmospheric composition of a planet can lead to insights into it's past (\citealt{Oberg2011}, \citealt{Madhu2014b}, \citealt{Mordasini2016}). One way of finding an atmosphere's composition is through consideration of the effects that a change in composition would have upon species in that atmosphere.

In this section we independently vary the \ce{C}/\ce{O} and \ce{N}/\ce{O} ratios for a planet otherwise equivalent to HD 189733b. We investigate the effects that composition has upon the abundance of a number of species in the atmosphere, how it affects the possibility of detecting some of these species and thus how easily detectable these changes in composition are. We explore a parameter space in which the total amount of carbon and nitrogen can vary between 0.1 and 10 times solar. This corresponds to an atomic \ce{C}/\ce{O} ratio varying between 0.055 and 5.5, and an \ce{N}/\ce{O} ratio varying between 0.0138 and 1.38. The atomic fraction of oxygen is always kept constant, with the change in carbon or nitrogen abundance being accounted for by an equivalent but opposite change in the total amount of hydrogen. We pick 100 mbar as the pressure being investigated since both emission and transmission spectra are sensitive to the abundance of species at this pressure. These models use the P-T and $K_{zz}$ profile of HD 189733b from Figures \ref{fig:PT} and \ref{fig:Kzzprof}.

\begin{figure}
   	\begin{subfigure}[t]{0.48\textwidth}
    	\centering
        \includegraphics[height=2.3in]{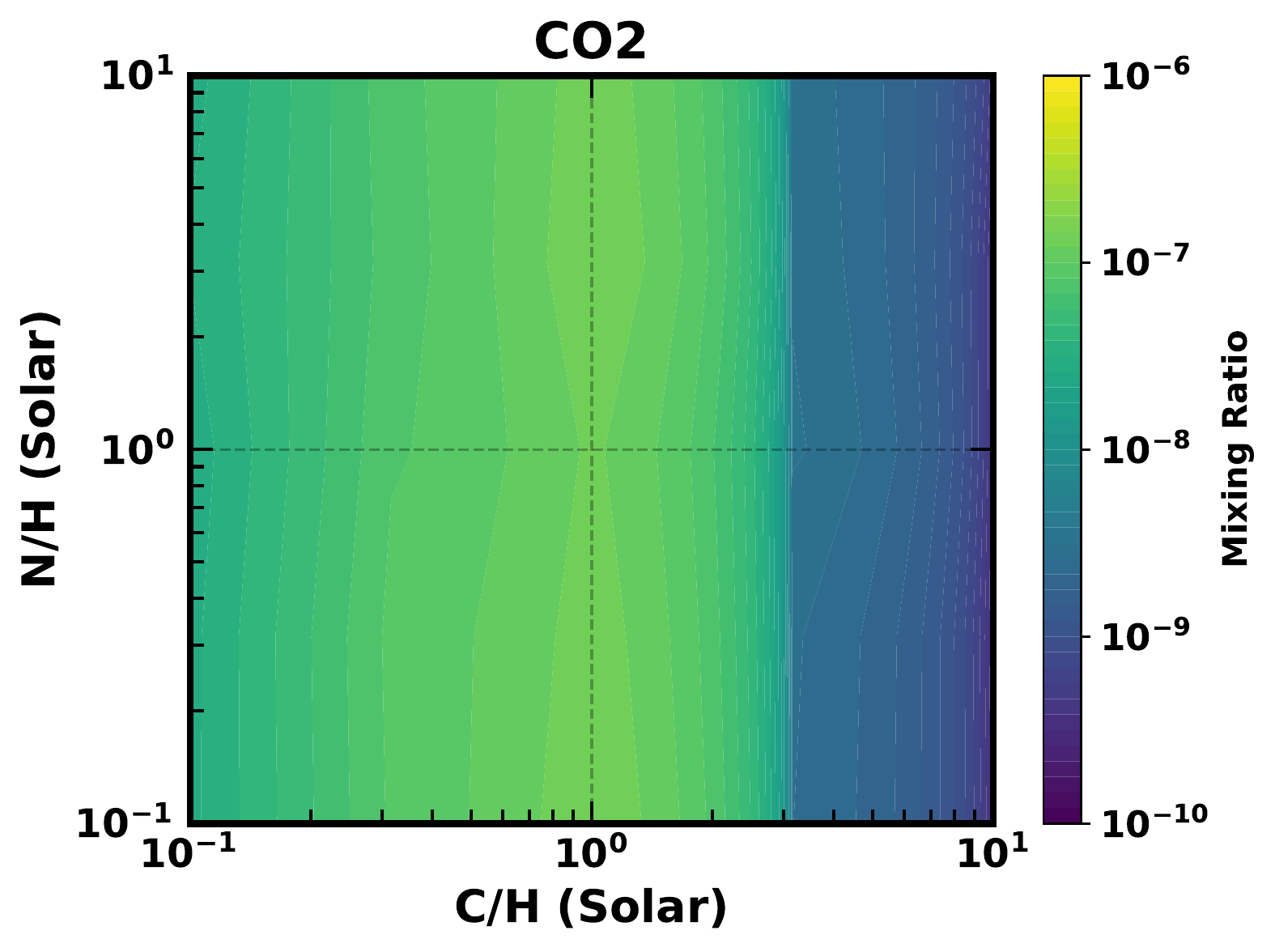}
        \caption{}
        \label{fig:ratioCO2}
	\end{subfigure}
    \begin{subfigure}[t]{0.48\textwidth}
    	\centering
        \includegraphics[height=2.3in]{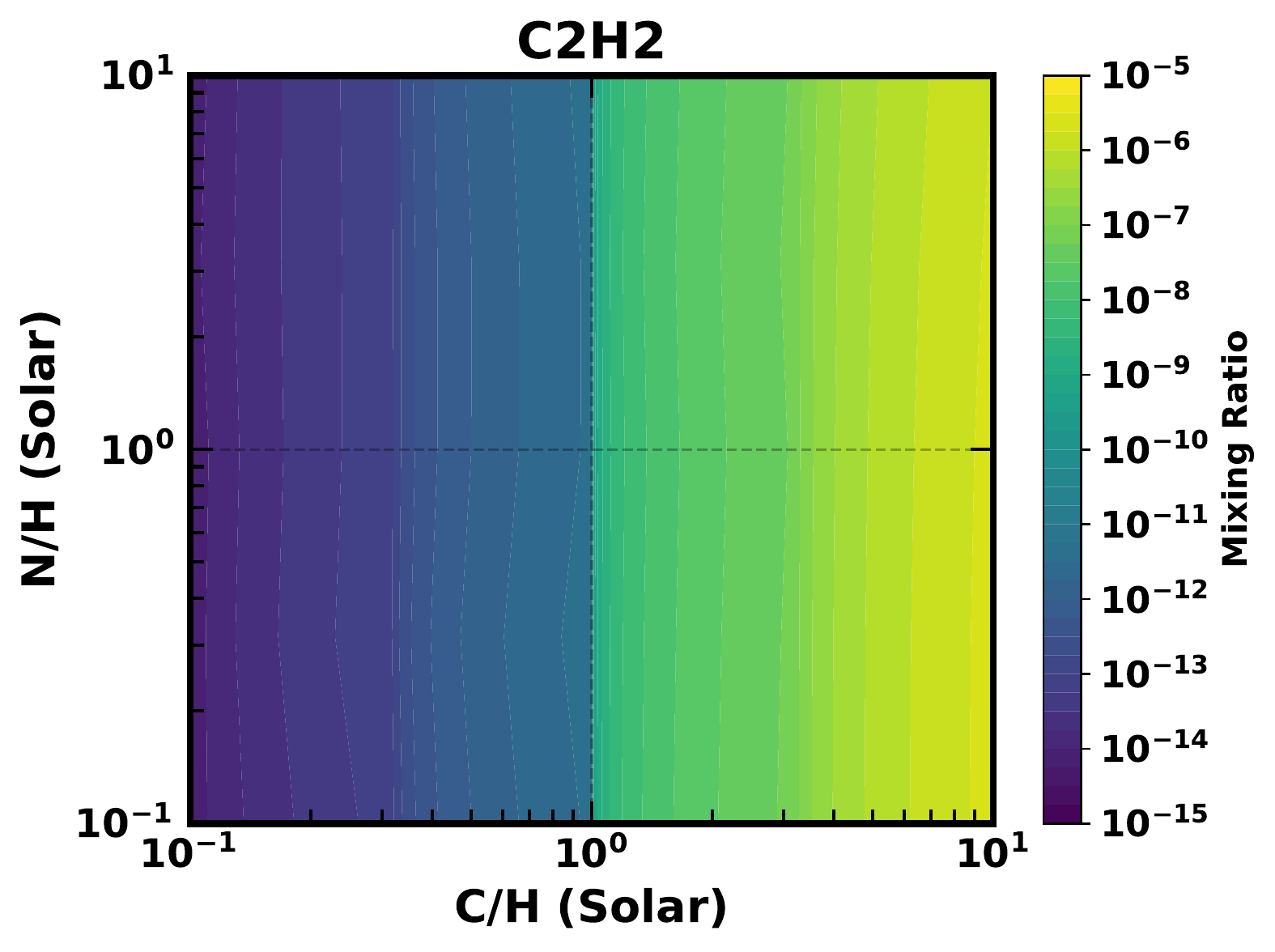}
        \caption{}
        \label{fig:ratioC2H2}
	\end{subfigure}
    \caption[Variation of carbon and oxygen species with varying C/N/O ratios2]{Abundance variations of \ce{CO2} (top), \ce{C2H2} (bottom) on HD 189733b at 0.1 bar for the parameter space in which the amount of nitrogen and carbon in the atmosphere varies between 0.1 and 10 times the solar amount. These models use the P-T and diffusion profile from Figure \ref{fig:PT} and \ref{fig:Kzzprof} for HD 189733b.}
    \label{figs:ratios1.5}
\end{figure}

In Figure \ref{figs:ratios1} and \ref{figs:ratios1.5} maps of the abundance of \ce{C} and \ce{O} species are shown as a function of atmospheric composition. None of the species show any strong dependence on the amount of \ce{N} in the atmosphere. In hydrogen dominated atmospheres, the abundances of \ce{H2O}, \ce{CH4} and \ce{CO} are primarily determined by the equation:
\begin{equation}
\ce{CH4 + H2O <=> CO + 3H2}
\label{eq:co}
\end{equation}
which at 100 mbar in the atmosphere of HD 189733b favours the formation of CO. At solar \ce{C}, a \ce{C}/\ce{O} ratio of 0.55, this means that \ce{CO} is the primary carbon carrier and \ce{H2O} contains most of the remaining oxygen. Since there is nearly twice as much oxygen as carbon in the atmosphere, \ce{CO} and \ce{H2O} have very similar abundances. For sub-solar carbon, there is less \ce{CO}, and thus less oxygen bound to \ce{CO}, resulting in an increase of \ce{H2O}. Once \ce{C}/\ce{O} > 1  (1.8$\times$ solar carbon), oxygen, not carbon, is now the limiting factor in producing \ce{CO}, and so the abundance of water quickly drops as there is very little oxygen available to form it. Much of the excess carbon ends up in \ce{CH4}, resulting in a sharp increase in its abundance. \ce{CO2} is also affected indirectly by equation \ref{eq:co}; while there is excess oxygen not bound to \ce{CO}, its abundance is directly related to the amount of carbon, however once \ce{C}/\ce{O} > 1, then the majority of the oxygen is bound to \ce{CO}, and the abundance of \ce{CO2} drops off rapidly. Larger hydrocarbons can also be very strongly affected by the variation in carbon, with the abundance of \ce{C2H2} changing by more than eight orders of magnitude for only two orders of magnitude change in carbon. 

\begin{figure}|
    \centering
    \begin{subfigure}[t]{0.48\textwidth}
    	\centering
        \includegraphics[height=2.3in]{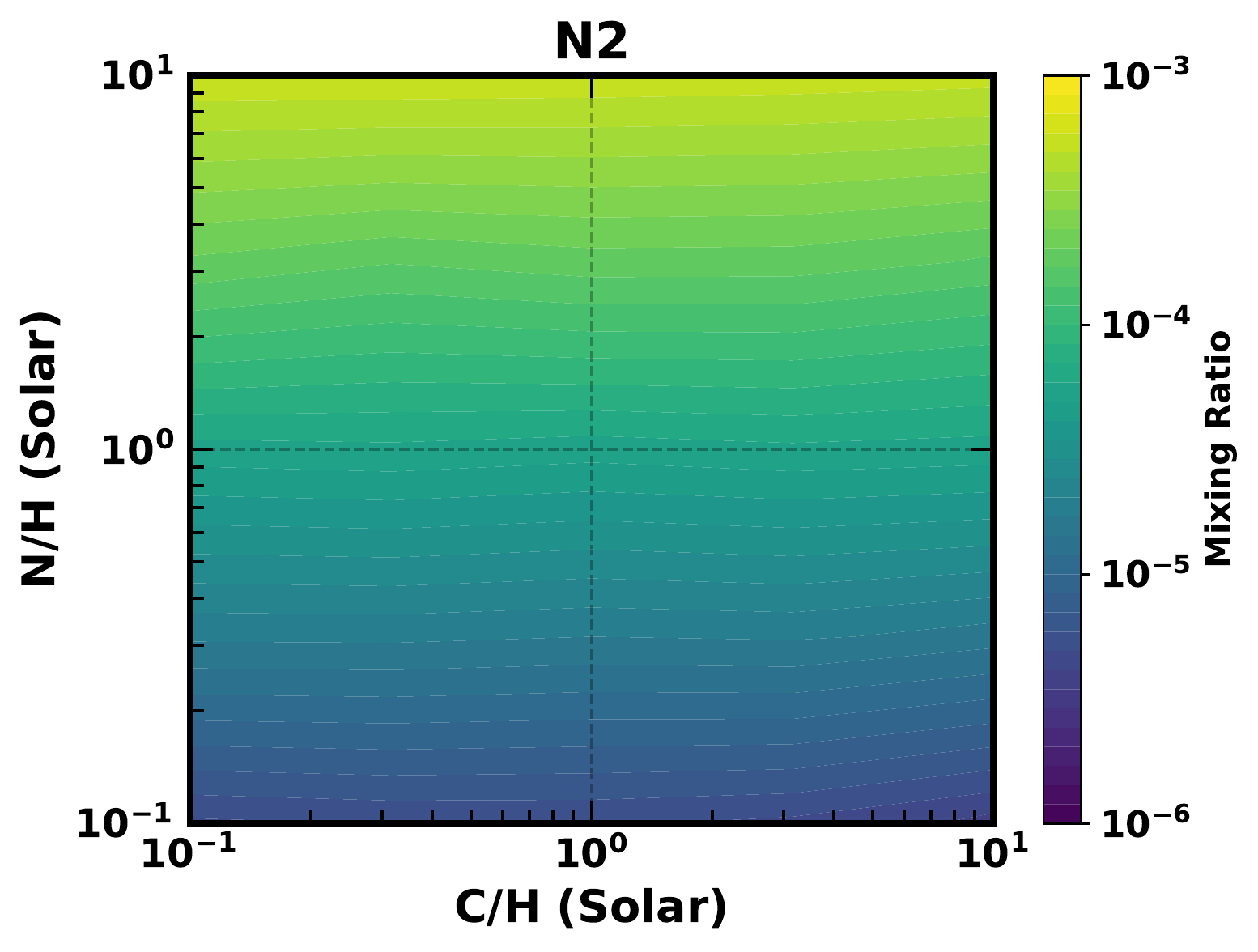}
        \caption{}
        \label{fig:ratioN2}
	\end{subfigure}
    \begin{subfigure}[t]{0.48\textwidth}
    	\centering
        \includegraphics[height=2.3in]{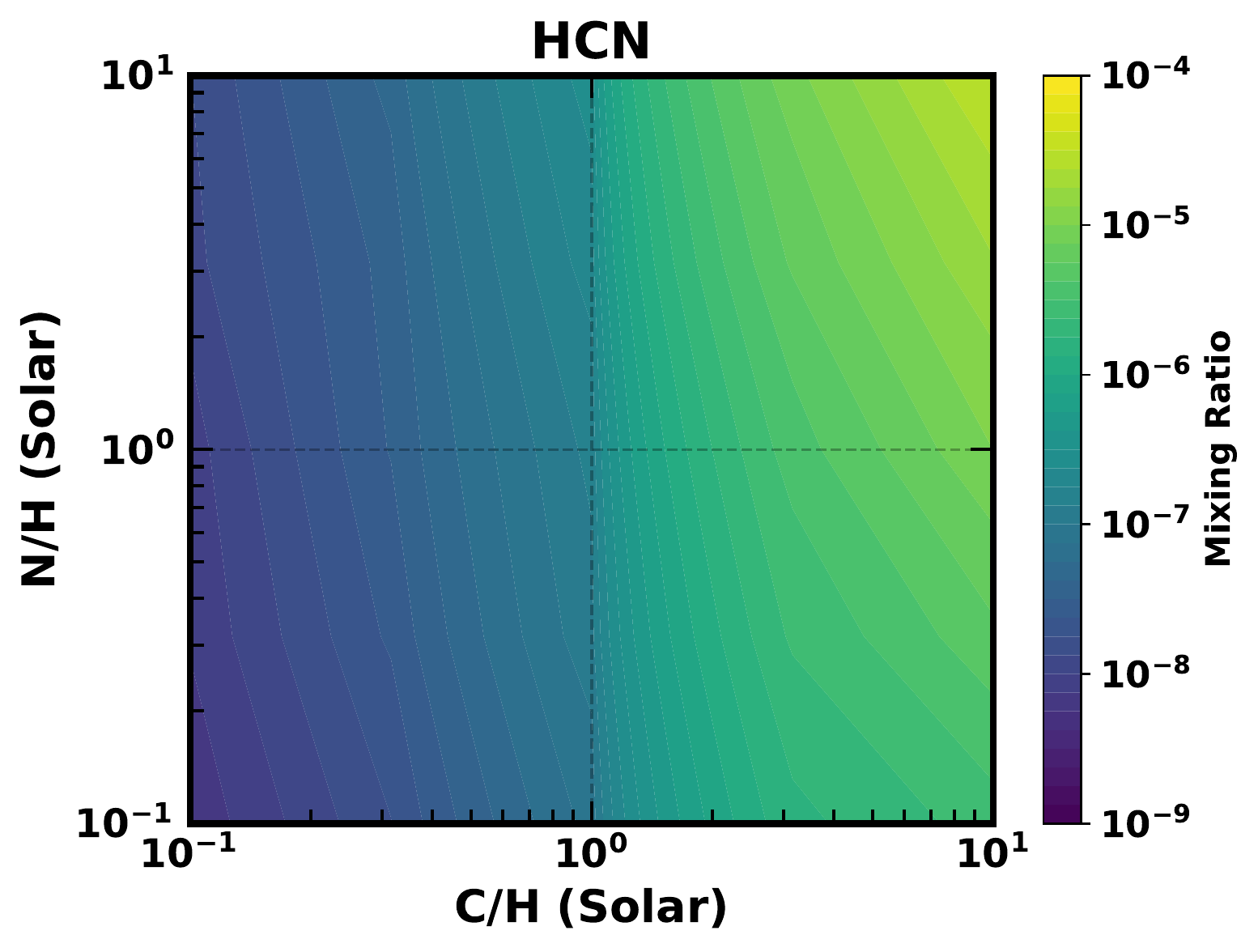}
        \caption{}
        \label{fig:RatioNH3}
	\end{subfigure}
    \begin{subfigure}[t]{0.48\textwidth}
    	\centering
        \includegraphics[height=2.3in]{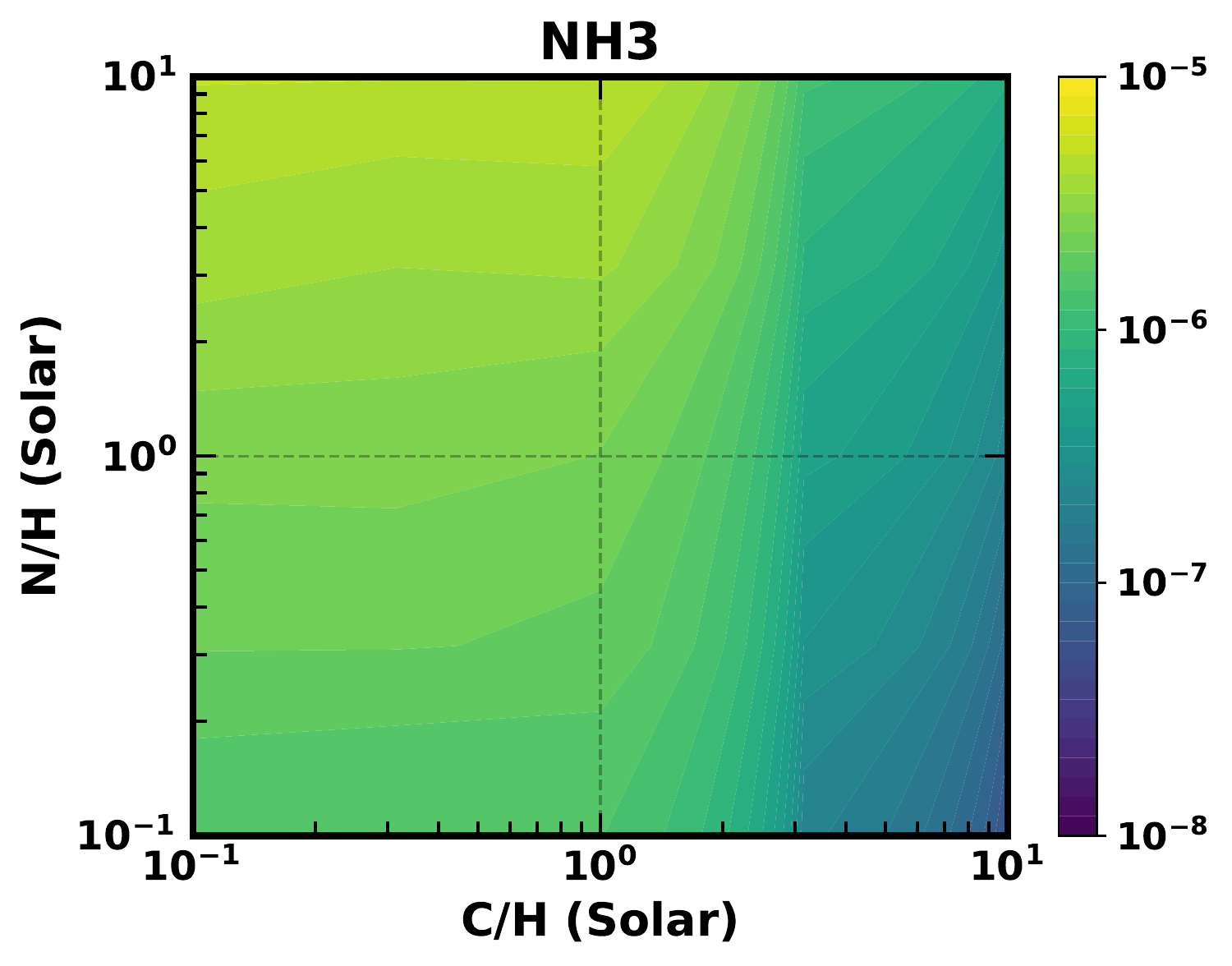}
        \caption{}
        \label{fig:ratioHCN}
	\end{subfigure}    
    \caption[Variation of nitrogen species with varying C/N/O ratios]{Abundance variations of \ce{N2} (top), \ce{NCN} (middle) and \ce{NH3} (bottom) on HD 189733b at 0.1 bar for the parameter space in which the amount of nitrogen and carbon in the atmosphere varies between 0.1 and 10 times the solar amount. These models use the P-T and diffusion profile from Figure \ref{fig:PT} and \ref{fig:Kzzprof} for HD 189733b.}
    \label{figs:ratios2}
\end{figure}

In Figure \ref{figs:ratios2} the variation of the three major nitrogen species within the parameter space is shown. The abundance of \ce{N2} has no dependence on carbon, while both \ce{NH3} and \ce{HCN} do. This is because of the fast reaction,
\begin{equation}
\begin{aligned}
\ce{H2 + M &-> H + H + M} \ (R5) \\
\ce{CH4 + H &-> CH3 + H2} \ (R251) \\
\ce{H + NH3 &-> NH2 + H2}  \ (R241) \\
\ce{H + NH2 &-> HN + H2}  \ (R220) \\
\ce{H + HN &-> N + H + H} \ (R204) \\
\ce{N + CH3 &-> HCN + H2} \ (R302) \\
\mathbf{Net:}\ \ce{NH3 + CH4 &-> HCN + 3H2},
\end{aligned}
\end{equation}
thus causing the abundance of \ce{NH3} and \ce{HCN} to have a dependence on the amount of carbon in the atmosphere. This dependence is particularly strong when \ce{C}/\ce{O} > 1, since there can now be an excess of \ce{CH4}, without it being converted into \ce{CO}. More \ce{CH4} therefore leads to more \ce{NH3} being converted into \ce{HCN}. As expected, all the nitrogen species have a strong dependence on the amount of nitrogen in the atmosphere, though \ce{HCN} has a weaker dependence than the others, due to carbon often being the limiting factor for its abundance. Many of these same trends for variation in the \ce{C}/\ce{O} ratio are seen in \cite{Madhu2012}. 

\begin{figure}
    \centering
    \begin{subfigure}[t]{0.48\textwidth}
    	\centering
        \includegraphics[height=2.3in]{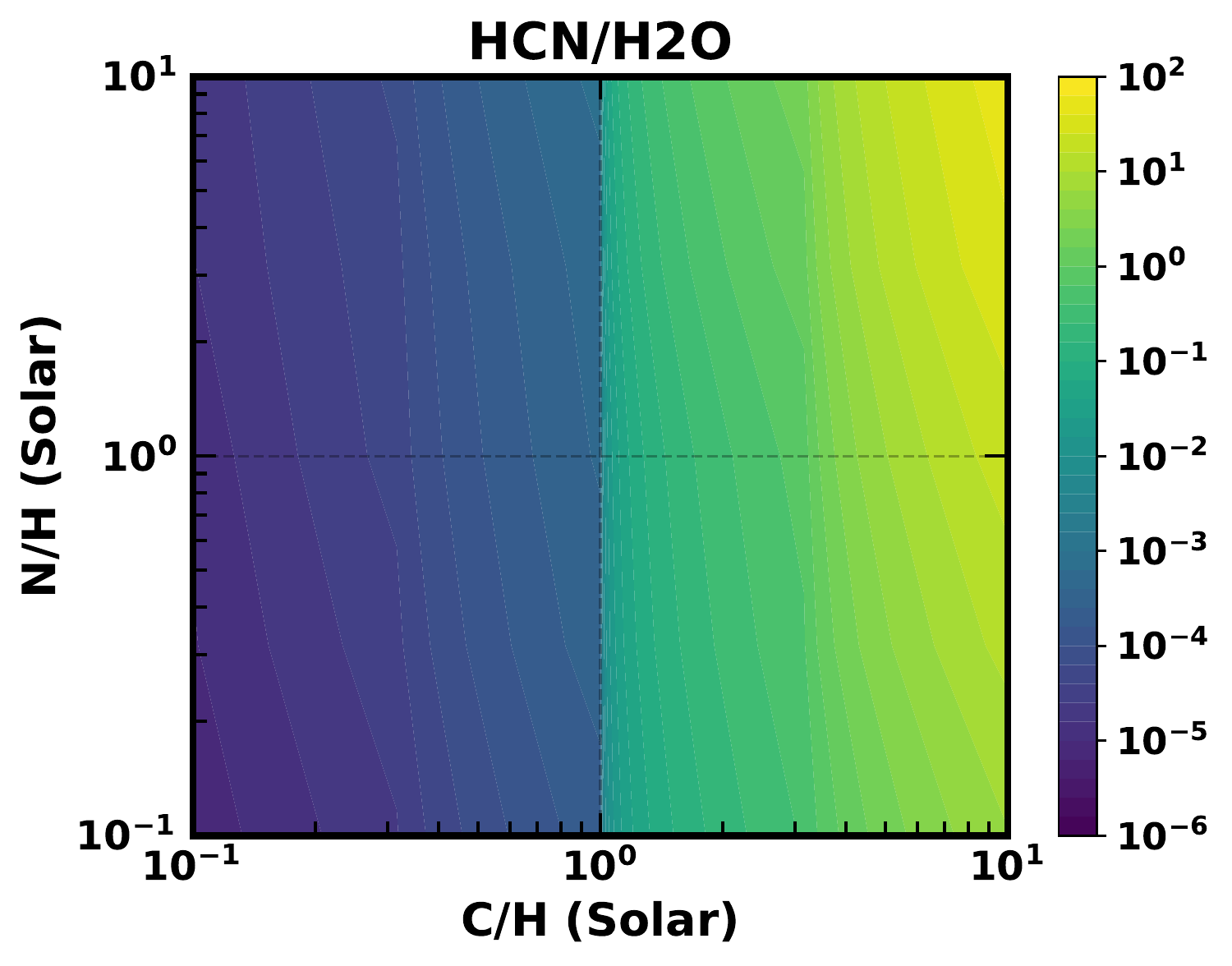}
        \caption{}
        \label{fig:ratioNH3H2O}
	\end{subfigure}
    \begin{subfigure}[t]{0.48\textwidth}
    	\centering
        \includegraphics[height=2.3in]{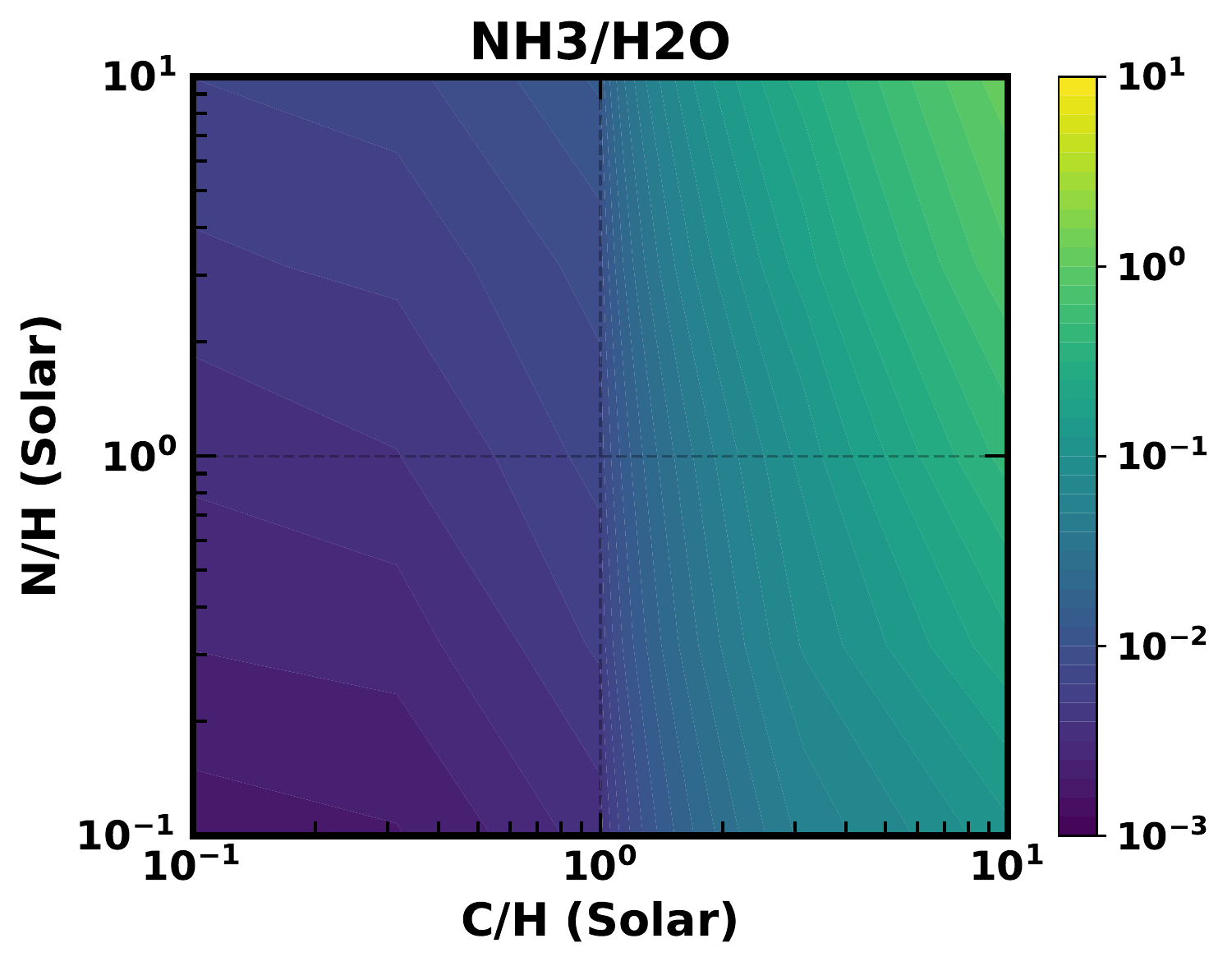}
        \caption{}
        \label{fig:RatioHCNH2O}
	\end{subfigure}   
    \caption[Observibility of nitrogen species with varying C/N/O ratios]{Variations in the ratio of \ce{HCN} and \ce{NH3} to \ce{H2O}, top and bottom respectively, on HD 189733b at 0.1 bar for the parameter space in which the amount of nitrogen and carbon in the atmosphere varies between 0.1 and 10 times the solar amount. These models use the P-T and diffusion profile from Figure \ref{fig:PT} and \ref{fig:Kzzprof} for HD 189733b.}
    \label{figs:ratiosratios}
\end{figure}

In Figure \ref{figs:ratiosratios}, the ratios of \ce{NH3} and \ce{HCN} to \ce{H2O} are shown, since this is what determines the observability of these species in exoplanet atmospheres. Since both \ce{NH3} and \ce{H2O} decrease with increasing carbon, the overall increase in \ce{NH3}/\ce{H2O} is due to the more rapid decrease of \ce{H2O}. Having \ce{HCN} and \ce{NH3} abundances at least 1\% that of \ce{H2O} is predicted to be needed to detect either of these molecules (\citealt{Macdonald2017}). Given this, Figure \ref{figs:ratiosratios} suggests that the amount of nitrogen is not very significant in the detection of these species, whereas \ce{C}/\ce{O} > 1 is almost always essential. However, exceptionally large amounts of nitrogen in the atmosphere can overcome this and allow detection of \ce{HCN} and \ce{NH3} at slightly lower \ce{C}/\ce{O} ratios.

\subsection{Molecular Detections on HD 209458b}
\label{sec:HD209again}

The atmosphere of HD 209458b was originally believed to contain a thermal inversion (\citealt{Knutson2008}), and so far in this work we have modelled it's atmosphere as such to enable validation of the code. More recent studies, however, have found that HD 209458b should not have a thermal inversion (\citealt{Diamond2014}), and it seems valuable for us to examine how this changes the predicted abundance profiles for species in its atmosphere. In addition, recent work in the literature presents evidence of \ce{CO}, \ce{H2O} and \ce{HCN} in the atmosphere of HD 209458b (\citealt{Hawker2018}). \cite{Hawker2018} find the best fit mixing abundance of \ce{HCN} to be $10^{-5}$, with a minimum mixing ratio of $10^{-6.5}$. They also find \ce{H2O} and \ce{CO} mixing ratios consistent with previous values of approximately $10^{-5\pm0.5}$ and $10^{-4\pm0.5}$ respectively (\citealt{Madhu2014,Macdonald2017a,Brogi2018}).

\begin{figure}
    \centering
    \includegraphics[width=0.5\textwidth]{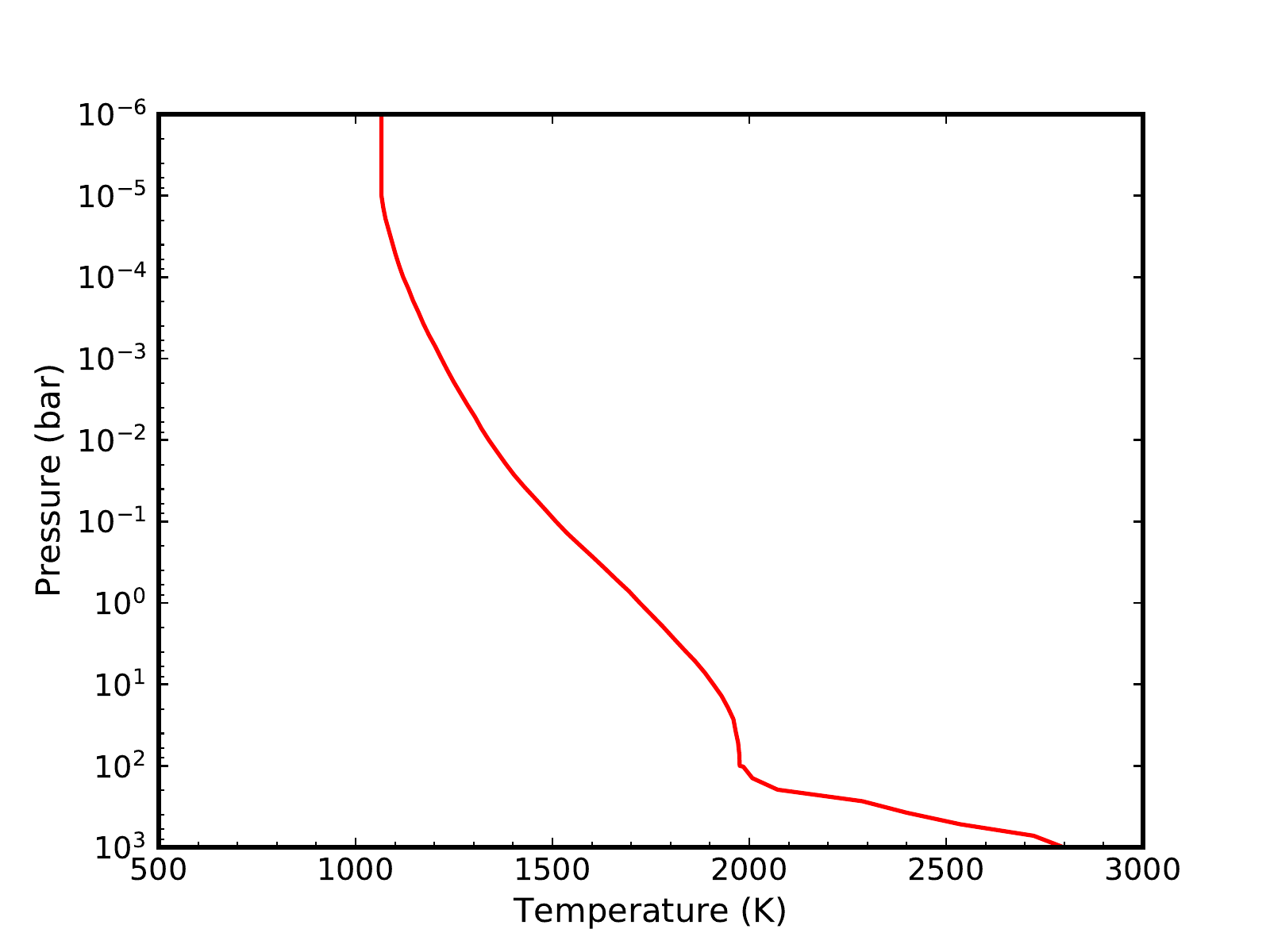}
    \caption[HD 290458b PT, No inversion]{The Pressure-Temperature profile for the average day-side of HD 209458b, with no thermal inversion in the atmosphere.}
    \label{fig:HD209NewPT}
\end{figure}

In this section we used a model P-T profile for the day-side of HD 209458b, without a thermal inversion (\citealt{Gandhi2017}). For the deep adiabatic region of the atmosphere, above 100 bars, we use the same values as the earlier P-T profile for HD 209458b, translated by -100 K for a smooth connection (Figure \ref{fig:HD209NewPT}). The same $K_{zz}$ profile and UV flux for HD 209458b from earlier are used, seen in Figure \ref{fig:Kzzprof} and Figure \ref{fig:Flux} respectively.

\begin{figure*}
    \centering
    \includegraphics{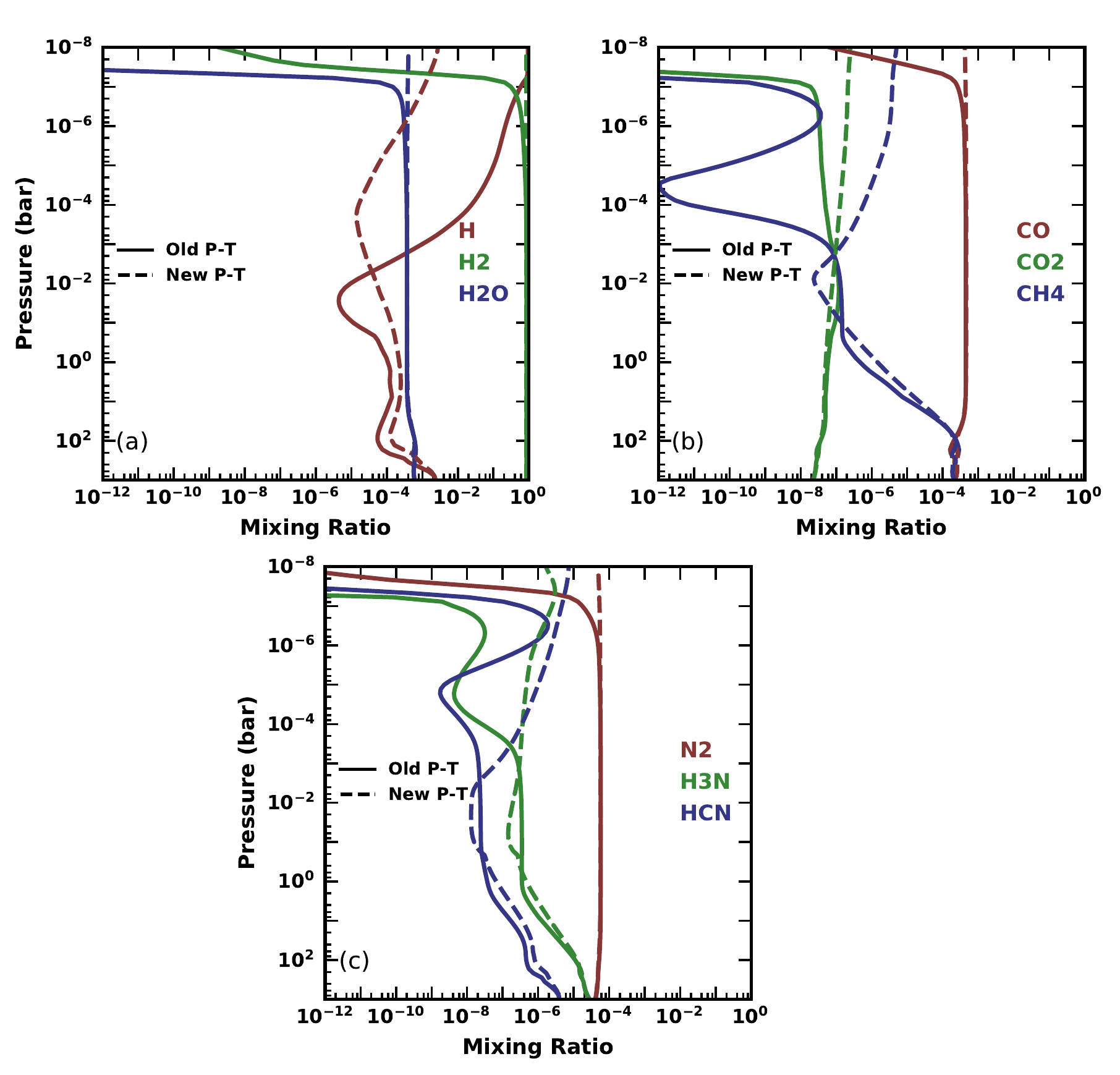}
    \caption[HD 290458b, New vs Old]{A comparison of the chemistry in the atmosphere of HD 209458b for two temperature profiles. The solid lines use a P-T profile with no thermal inversion, from Figure \ref{fig:HD209NewPT}, while the dashed lines use a P-T profile with a thermal inversion, from Figure \ref{fig:PT}}
    \label{fig:HD209NewOld}
\end{figure*}

In Figure \ref{fig:HD209NewOld} the chemical abundance profiles for the new P-T profile of HD 209458b are compared with those from the old P-T profile. In the lower atmosphere, where the P-T profile is largely unchanged, there is very little difference between the old and new mixing ratios. In the upper atmosphere the lower temperature of the new P-T profile, due to the lack of a thermal inversion, means that some species that would otherwise be depleted by the higher temperatures, such as \ce{HCN} or \ce{CH4}, are present in greater quantities. For \ce{HCN} and \ce{CH4} this results in an increase in abundance of nearly two orders of magnitude at $10^{-3}$ bar. 

\begin{figure*}
    \centering
    \includegraphics{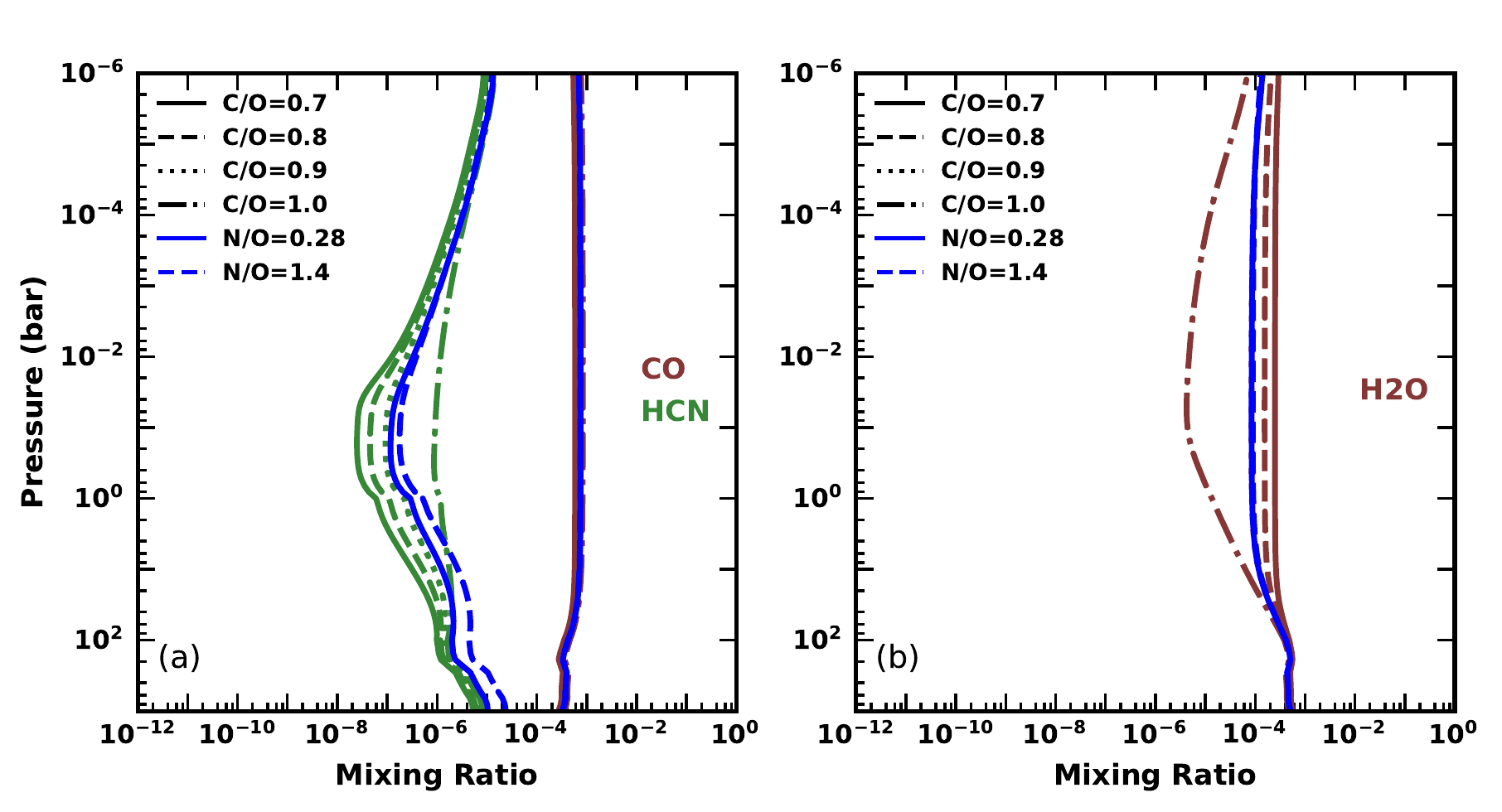}
    \caption[HD 290458b Atmosphere, No inversion]{The mixing ratios of \ce{CO}, \ce{HCN} and \ce{H2O} on HD 209458b, using the P-T profile from Figure \ref{fig:HD209NewPT}. Each of the molecules are displayed at different \ce{C}/\ce{O} and \ce{N}/\ce{O} ratios. The \ce{C}/\ce{O} ratio is 0.9 for the two \ce{N}/\ce{O} ratios shown.}
    \label{fig:HD209NewCO}
\end{figure*}

In Figure \ref{fig:HD209NewCO} the abundance profiles for \ce{CO}, \ce{HCN} and \ce{H2O} are displayed for a range of \ce{C}/\ce{O} ratios.
We can compare the mixing ratios of these species between $10^{-1}$ and $10^{-3}$ bar in our models to the estimated abundances from \cite{Hawker2018}. The abundance of \ce{CO} provides little information since across all \ce{C}/\ce{O} ratios being modelled it stays at approximately $10^{-3}$, consistent with the expected value. \ce{HCN} provides more information, with a \ce{C}/\ce{O} $\gtrsim$ 0.9 required to pass the minimum estimated mixing ratio of $10^{-6.5}$, and a \ce{C}/\ce{O} = 1.2 is required for \ce{HCN} to match the best fit abundance of $10^{-5}$. However, \ce{H2O} provides an upper limit of \ce{C}/\ce{O} $\approx$ 1; higher \ce{C}/\ce{O} than this quickly depletes \ce{H2O} below $10^{-6}$. As such, it is clear that alterations to the \ce{C}/\ce{O} ratio by itself may not be sufficient to explain the observed abundances. 

In Figure \ref{fig:HD209NewCO} we also included two models in which we increased the \ce{N}/\ce{O} ratio by two times (\ce{N}/\ce{O} = 0.28) and ten times (\ce{N}/\ce{O}=1.4) the solar value, to discover the effect this has on the abundance of these molecules. For these models, we kept \ce{C}/\ce{O} = 0.9. Increasing the amount of nitrogen in the atmosphere by two times solar increases the \ce{HCN} abundance by approximately a third at $10^{-3}$ bar. Further increases to the amount of nitrogen in the atmosphere do little to the abundance of \ce{HCN} since the amount of free carbon is the limiting factor beyond this. The \ce{N}/\ce{O} ratio has no significant effect on the abundance of either \ce{CO} or \ce{H2O}.

Through comparison of our model and the expected molecular abundances of HD 209458b we can produce some initial constraints for the \ce{C}/\ce{O} and \ce{N}/\ce{O} ratio of this hot Jupiter: a 0.9 $\lesssim$ \ce{C}/\ce{O} $\gtrsim$ 1, and preferentially \ce{N}/\ce{O} > 1. However, varying \ce{C}/\ce{O} and \ce{N}/\ce{O} is not sufficient to match the estimated best fit abundance of \ce{HCN}, which lies outside the range that these two parameters alone can explore. In particular, there are a number of other important parameters that could and should also be varied, such as the $K_{zz}$ strength, the atmospheric metallicity and the P-T profile. This full sweep of parameter space is, however, beyond the scope of this work, but provides a good basis for future investigations.
 
\section{Conclusion} 

\label{sec:Futurework} 

In this work, we present a new one-dimensional diffusion and photochemistry code, named \textsc{Levi}, currently able to model gaseous chemistry in hot Jupiter atmospheres via solving of the continuity equation. We focus on \ce{H}, \ce{C}, \ce{O} and \ce{N} chemistry, using a chemical network that contains over 1000 reactions and 150 species. Through inputs in the form of a pressure - temperature profile, an eddy diffusion profile and a UV stellar flux, the code can calculate steady state abundance profiles for all species over the desired pressure range.

\text{Levi} was validated against the disequilibrium chemistry models produced by \cite{Moses2011} and \cite{Venot2012}; For HD 209458b the abundance profiles of species that \text{Levi} produced matched closely with those of the other two models, with any differences being a consequence of the slightly differing input parameters. We discussed how the differing P-T profiles of HD 209458b and HD 189733b affected equilibrium chemistry in their atmosphere, and noted that, in chemical equilibrium, neither \ce{HCN} nor \ce{NH3} would be detectable. Eddy-diffusion and photochemistry were also included, and we discussed how they affected the chemistry of the atmospheres, as well as how the chemistry of the two planets atmospheres compared to each other. Of particular note was that the quenching caused by eddy-diffusion is capable of raising the abundance of \ce{NH3} and \ce{HCN} to potentially detectable levels.

The influence of the parameter space of the \ce{C}/\ce{O} and \ce{N}/\ce{O} ratio on the abundance of various molecules at 100 mbar in the atmosphere of HD 189733b was investigated. It was found that species that contained \ce{C} or \ce{O} were strongly affected by variations in the \ce{C}/\ce{O} ratio, while other molecules were weakly affected, if at all. In general, only species that contain nitrogen are affected by changes to the \ce{N}/\ce{O} ratio. We also looked at how the fractions \ce{NH3}/\ce{H2O} and \ce{HCN}/\ce{H2O} varied with the atmospheric composition, an important fraction in determining whether it is possible to detect these molecules. We found that when trying to detect \ce{NH3} or \ce{HCN}, the quantity of carbon in the atmosphere is much more important than the amount of nitrogen. At a solar \ce{N}/\ce{O} ratio, sub-solar carbon would make detecting either of these molecules near impossible, however with at least 5 times solar nitrogen in the atmosphere, \ce{NH3} and \ce{HCN} could be detected with only 0.5 times and 0.8 times solar carbon respectively. In an atmosphere with a deficit of nitrogen, less than two times solar carbon would make detecting either of these molecules possible.

Recent literature (\citealt{Diamond2014}) has suggested that HD 209458b does not contain a thermal inversion. Other work (\citealt{Madhu2014,Macdonald2017a,Brogi2018,Hawker2018}) has found evidence for \ce{CO}, \ce{H2O} and \ce{HCN} with best fit mixing ratios of approximately $10^{-4\pm0.5}$, $10^{-5\pm0.5} $ and $10^{-5}$ respectively, and with a minimum mixing ratio of \ce{HCN} at $10^{-6.5}$. We applied our model to HD 209458b, using a new P-T profile without a thermal inversion, to discover which \ce{C}/\ce{O} and \ce{N}/\ce{O} ratios could best fit with these observations. A \ce{C}/\ce{O} > 1 resulted in a significant depletion of \ce{H2O}. To obtain at least the minimum mixing ratio of \ce{HCN}, a \ce{C}/\ce{O} $\gtrsim$ 0.9 was required. A large \ce{N}/\ce{O} ratio, ten times solar or more, can help increase the \ce{HCN} abundance closer to the best fit value, but it results in much smaller increases compared to increasing the \ce{C}/\ce{O} ratio. Further testing with both these parameters and others are outside the scope of this work, and have been left for future investigations.

\section*{Acknowledgements}

R.H. and N.M. acknowledge support from the UK Science and Technology Facilities Council (STFC). PBR acknowledges funding from the Simons Foundation SCOL awards 599634. 




\bibliographystyle{mnras}
\bibliography{bibliography}






\bsp	
\label{lastpage}
\end{document}